\documentclass[aps,prl,reprint,superscriptaddress,twocolumn]{revtex4-1}

\usepackage{graphicx}
\usepackage{amsmath}
\usepackage{amssymb}
\usepackage{units}
\usepackage{color}
\usepackage{hyperref}

\newcommand{\ssr}{Space Science Reviews}

\begin{document}

\title{Observation of High-Energy Astrophysical Neutrinos in Three Years of IceCube Data}

\affiliation{III. Physikalisches Institut, RWTH Aachen University, D-52056 Aachen, Germany}
\affiliation{School of Chemistry \& Physics, University of Adelaide, Adelaide SA, 5005 Australia}
\affiliation{Dept.~of Physics and Astronomy, University of Alaska Anchorage, 3211 Providence Dr., Anchorage, AK 99508, USA}
\affiliation{CTSPS, Clark-Atlanta University, Atlanta, GA 30314, USA}
\affiliation{School of Physics and Center for Relativistic Astrophysics, Georgia Institute of Technology, Atlanta, GA 30332, USA}
\affiliation{Dept.~of Physics, Southern University, Baton Rouge, LA 70813, USA}
\affiliation{Dept.~of Physics, University of California, Berkeley, CA 94720, USA}
\affiliation{Lawrence Berkeley National Laboratory, Berkeley, CA 94720, USA}
\affiliation{Institut f\"ur Physik, Humboldt-Universit\"at zu Berlin, D-12489 Berlin, Germany}
\affiliation{Fakult\"at f\"ur Physik \& Astronomie, Ruhr-Universit\"at Bochum, D-44780 Bochum, Germany}
\affiliation{Physikalisches Institut, Universit\"at Bonn, Nussallee 12, D-53115 Bonn, Germany}
\affiliation{Universit\'e Libre de Bruxelles, Science Faculty CP230, B-1050 Brussels, Belgium}
\affiliation{Vrije Universiteit Brussel, Dienst ELEM, B-1050 Brussels, Belgium}
\affiliation{Dept.~of Physics, Chiba University, Chiba 263-8522, Japan}
\affiliation{Dept.~of Physics and Astronomy, University of Canterbury, Private Bag 4800, Christchurch, New Zealand}
\affiliation{Dept.~of Physics, University of Maryland, College Park, MD 20742, USA}
\affiliation{Dept.~of Physics and Center for Cosmology and Astro-Particle Physics, Ohio State University, Columbus, OH 43210, USA}
\affiliation{Dept.~of Astronomy, Ohio State University, Columbus, OH 43210, USA}
\affiliation{Niels Bohr Institute, University of Copenhagen, DK-2100 Copenhagen, Denmark}
\affiliation{Dept.~of Physics, TU Dortmund University, D-44221 Dortmund, Germany}
\affiliation{Dept.~of Physics, University of Alberta, Edmonton, Alberta, Canada T6G 2E1}
\affiliation{Erlangen Centre for Astroparticle Physics, Friedrich-Alexander-Universit\"at Erlangen-N\"urnberg, D-91058 Erlangen, Germany}
\affiliation{D\'epartement de physique nucl\'eaire et corpusculaire, Universit\'e de Gen\`eve, CH-1211 Gen\`eve, Switzerland}
\affiliation{Dept.~of Physics and Astronomy, University of Gent, B-9000 Gent, Belgium}
\affiliation{Dept.~of Physics and Astronomy, University of California, Irvine, CA 92697, USA}
\affiliation{Dept.~of Physics and Astronomy, University of Kansas, Lawrence, KS 66045, USA}
\affiliation{Dept.~of Astronomy, University of Wisconsin, Madison, WI 53706, USA}
\affiliation{Dept.~of Physics and Wisconsin IceCube Particle Astrophysics Center, University of Wisconsin, Madison, WI 53706, USA}
\affiliation{Institute of Physics, University of Mainz, Staudinger Weg 7, D-55099 Mainz, Germany}
\affiliation{Universit\'e de Mons, 7000 Mons, Belgium}
\affiliation{T.U. Munich, D-85748 Garching, Germany}
\affiliation{Bartol Research Institute and Dept.~of Physics and Astronomy, University of Delaware, Newark, DE 19716, USA}
\affiliation{Dept.~of Physics, University of Oxford, 1 Keble Road, Oxford OX1 3NP, UK}
\affiliation{Dept.~of Physics, South Dakota School of Mines and Technology, Rapid City, SD 57701, USA}
\affiliation{Dept.~of Physics, University of Wisconsin, River Falls, WI 54022, USA}
\affiliation{Oskar Klein Centre and Dept.~of Physics, Stockholm University, SE-10691 Stockholm, Sweden}
\affiliation{Dept.~of Physics and Astronomy, Stony Brook University, Stony Brook, NY 11794-3800, USA}
\affiliation{Dept.~of Physics, Sungkyunkwan University, Suwon 440-746, Korea}
\affiliation{Dept.~of Physics, University of Toronto, Toronto, Ontario, Canada, M5S 1A7}
\affiliation{Dept.~of Physics and Astronomy, University of Alabama, Tuscaloosa, AL 35487, USA}
\affiliation{Dept.~of Astronomy and Astrophysics, Pennsylvania State University, University Park, PA 16802, USA}
\affiliation{Dept.~of Physics, Pennsylvania State University, University Park, PA 16802, USA}
\affiliation{Dept.~of Physics and Astronomy, Uppsala University, Box 516, S-75120 Uppsala, Sweden}
\affiliation{Dept.~of Physics, University of Wuppertal, D-42119 Wuppertal, Germany}
\affiliation{DESY, D-15735 Zeuthen, Germany}

\author{M.~G.~Aartsen}
\affiliation{School of Chemistry \& Physics, University of Adelaide, Adelaide SA, 5005 Australia}
\author{M.~Ackermann}
\affiliation{DESY, D-15735 Zeuthen, Germany}
\author{J.~Adams}
\affiliation{Dept.~of Physics and Astronomy, University of Canterbury, Private Bag 4800, Christchurch, New Zealand}
\author{J.~A.~Aguilar}
\affiliation{D\'epartement de physique nucl\'eaire et corpusculaire, Universit\'e de Gen\`eve, CH-1211 Gen\`eve, Switzerland}
\author{M.~Ahlers}
\affiliation{Dept.~of Physics and Wisconsin IceCube Particle Astrophysics Center, University of Wisconsin, Madison, WI 53706, USA}
\author{M.~Ahrens}
\affiliation{Oskar Klein Centre and Dept.~of Physics, Stockholm University, SE-10691 Stockholm, Sweden}
\author{D.~Altmann}
\affiliation{Erlangen Centre for Astroparticle Physics, Friedrich-Alexander-Universit\"at Erlangen-N\"urnberg, D-91058 Erlangen, Germany}
\author{T.~Anderson}
\affiliation{Dept.~of Physics, Pennsylvania State University, University Park, PA 16802, USA}
\author{C.~Arguelles}
\affiliation{Dept.~of Physics and Wisconsin IceCube Particle Astrophysics Center, University of Wisconsin, Madison, WI 53706, USA}
\author{T.~C.~Arlen}
\affiliation{Dept.~of Physics, Pennsylvania State University, University Park, PA 16802, USA}
% ...

\collaboration{IceCube Collaboration}
\noaffiliation

\begin{abstract}
A search for high-energy neutrinos interacting within the IceCube detector between 2010 and 2012 provided the first evidence for a high-energy neutrino flux of extraterrestrial origin. Results from an analysis using the same methods with a third year (2012-2013) of data from the complete IceCube detector are consistent with the previously reported astrophysical flux in the 100 TeV - PeV range at the level of $10^{-8}\, \mathrm{GeV}\, \mathrm{cm}^{-2}\, \mathrm{s}^{-1}\, \mathrm{sr}^{-1}$ per flavor and reject a purely atmospheric explanation for the combined 3-year data at $5.7 \sigma$. The data are consistent with expectations for equal fluxes of all three neutrino flavors and with isotropic arrival directions, suggesting either numerous or spatially extended sources. The three-year data set, with a livetime of 988 days, contains a total of 37 neutrino candidate events with deposited energies ranging from 30 to 2000 TeV. The 2000 TeV event is the highest-energy neutrino interaction ever observed.
\end{abstract}

\maketitle


High energy neutrinos are expected to be produced in astrophysical objects by the decays of charged pions made in cosmic ray interactions with radiation or gas \cite{1978ApJ...221..990M,1979ApJ...228..919S,1989cgrc.conf...21B,1990JPhG...16.1917M}. As these pions decay, they produce neutrinos with typical energies of 5\% those of the cosmic ray nucleons \cite{Mucke:1999yb,Kelner:2006tc}. These neutrinos can travel long distances undisturbed by either the absorption experienced by high-energy photons or the magnetic deflection experienced by charged particles, making them a unique tracer of cosmic ray acceleration.

Observations since 2008 using the Antarctic gigaton IceCube detector \cite{daqpaper} while it was under construction provided several indications of such neutrinos in a variety of channels \cite{ic40_cascades,ic59_muons,ehepaper}. Two years of data from the full detector, from May 2010 - May 2012, then provided the first strong evidence for the detection of these astrophysical neutrinos \cite{hese_paper} using an all-flavor all-direction sample of neutrinos interacting within the detector volume. This analysis focused on neutrinos above 100 TeV, at which the expected atmospheric neutrino background falls to the level of one event per year, allowing any harder astrophysical flux to be seen clearly. Here, following the same techniques, we add a third year of data supporting this result and begin to probe the properties of the observed astrophysical neutrino flux.

Neutrinos are detected in IceCube by observing the Cherenkov light produced in ice by charged particles created when neutrinos interact. These particles generally travel distances too small to be resolved individually and the particle shower is observed only in aggregate. In $\nu_\mu$ charged-current (CC) interactions, however, as well as a minority of $\nu_\tau$ CC, a high-energy muon is produced that leaves a visible track (unless produced on the detector boundary heading outward). Although deposited energy resolution is similar for all events, angular resolution for events containing visible muon tracks is much better ($\lesssim 1^\circ$, 50\% CL) than for those that do not ($\sim 15^\circ$, 50\% CL) \cite{energy_reco}.  For equal neutrino fluxes of all flavors (1:1:1), $\nu_\mu$ CC events make up only 20\% of interactions \cite{2004JCAP...11..009B}.


Backgrounds to astrophysical neutrino detection arise entirely from cosmic ray air showers. Muons produced by $\pi$ and $K$ decays above IceCube enter the detector at 2.8 kHz. Neutrinos produced in the same interactions \cite{1983PhRvL..51..223G,2006JHEP...10..075G,2007PhRvD..75l3005L,2012PhRvD..86k4024F} enter IceCube from above and below, and are seen at a much lower rate due to the low neutrino interaction cross-section. Because $\pi$ and $K$ mesons decay overwhelmingly to muons rather than electrons, these neutrinos are predominantly $\nu_\mu$ and usually have track-type topologies in the detector \cite{2004JCAP...11..009B}. As the parent meson's energy rises, its lifetime increases, making it increasingly likely to interact before decaying. Both the atmospheric muon and neutrino fluxes thus become suppressed at high energy, with a spectrum one power steeper than the primary cosmic rays that produced them \cite{Honda2006}. At energies above $\sim100$ TeV, an analogous flux of muons and neutrinos from the decay of charmed mesons is expected to dominate, as the shorter lifetime of these particles allows this flux to avoid suppression from interaction before decay \cite{1978PhLB...78..635B,1983ICRC....7...22V,1999PhLB..462..211V,2000PhRvD..61e6011G,2003AcPPB..34.3273M,2008PhRvD..78d3005E,2013AIPC.1560..350R}. This flux has not yet been observed, however, and both its overall rate and cross-over energy with the $\pi/K$ flux are at present poorly constrained \cite{charm}. As before \cite{hese_paper}, we estimate all atmospheric neutrino background rates using measurements of the northern-hemisphere $\nu_\mu$ spectrum \cite{ic59_muons}.

Event selection identifies neutrino interactions in IceCube by rejecting those events with Cherenkov-radiating particles, principally cosmic ray muons, entering from outside the detector. As before, we used a simple anticoincidence muon veto in the outer layers of the detector \cite{hese_paper}, requiring that fewer than 3 of the first 250 detected photoelectrons (PE) be on the detector boundary. To ensure sufficient numbers of photons to reliably trigger this veto, we additionally required at least 6000 PE overall, corresponding to deposited energies of approximately 30 TeV. This rejects all but one part in $10^5$ of the cosmic ray muon background above 6000 PE while providing a direction and topology-neutral neutrino sample \cite{hese_paper}. We use a data-driven method to estimate this background by using one region of IceCube to tag muons and then measuring their detection rate in a separate layer of PMTs equivalent to our veto; this predicts a total muon background in three years of $8.4 \pm 4.2$ events. Rejection of events containing entering muons also significantly reduces downgoing atmospheric neutrinos (the southern hemisphere) by detecting and vetoing muons produced in the neutrinos' parent air showers \cite{atmonu_veto,newvetopaper}. This southern-hemisphere suppression is a distinctive and generic feature of any neutrinos originating in cosmic ray interactions in the atmosphere.


\begin{figure}
\includegraphics[width=\linewidth]{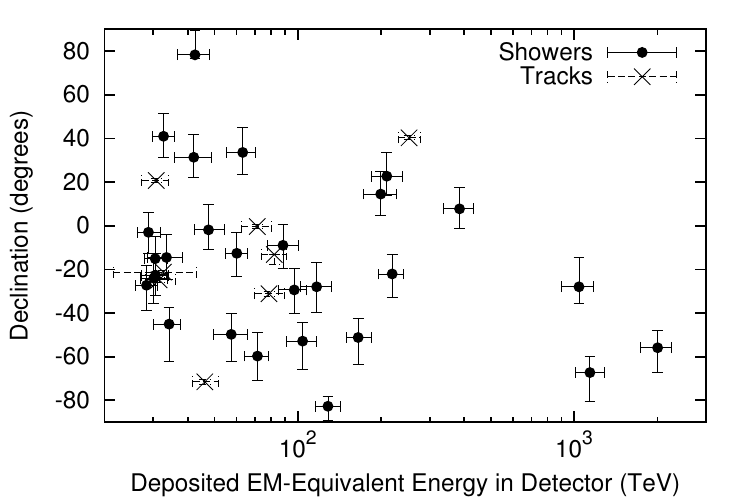}
\caption{Arrival angles and deposited energies of the events. Cosmic ray muon background would appear as low-energy track events in the southern sky (bottom). Atmospheric neutrino backgrounds would appear primarily in the northern sky (top), also at low energies and predominantly as tracks. The attenuation of high energy neutrinos in the Earth is visible in the top right of the figure. One event, a pair of coincident unrelated cosmic ray muons, is excluded from this plot. A tabular version of these data, including additional information such as event times, can be found in the online supplement~\cite{supplemental_section}.}
\label{fig:energyzenith}
\end{figure}

In the full 988-day sample, we detected 37 events (Fig.~\ref{fig:energyzenith}) with these characteristics relative to an expected background of $8.4 \pm 4.2$ cosmic ray muon events and $6.6 ^{+5.9} _{-1.6}$ atmospheric neutrinos. Nine were observed in the third year.  One of these (event 32) was produced by a coincident pair of background muons from unrelated air showers. This event cannot be reconstructed with a single direction and energy and is excluded from the remainder of this article where these quantities are required. This event, like event 28, had sub-threshold early hits in the IceTop surface array and our veto region, and is likely part of the expected muon background. Three additional downgoing track events are ambiguous; the remainder are uniformly distributed through the detector and appear to be neutrino interactions.

\begin{figure}
\includegraphics[width=\linewidth]{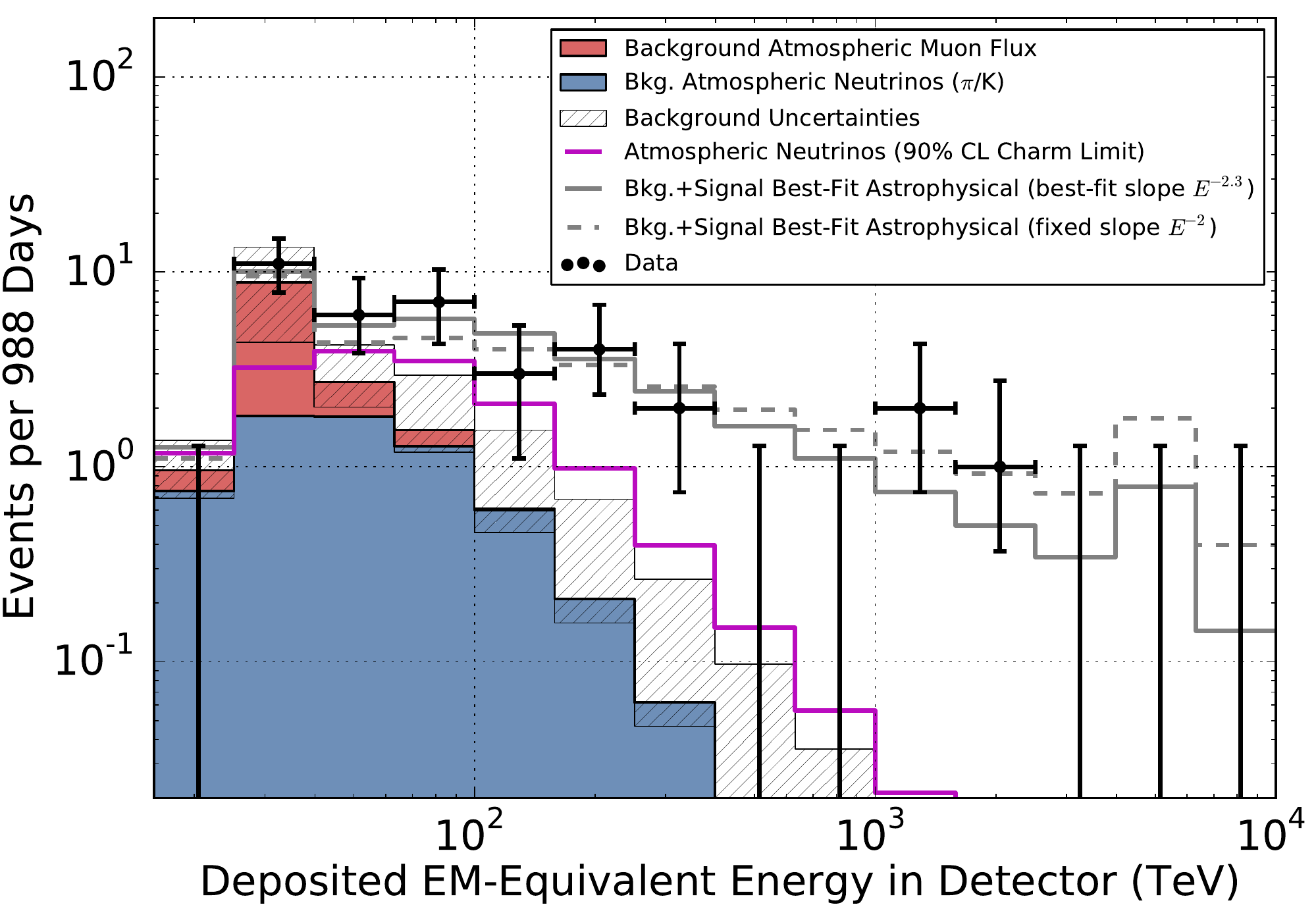}
\caption{Deposited energies of observed events with predictions. The hashed region shows uncertainties on the sum of all backgrounds. Muons (red) are computed from simulation to overcome statistical limitations in our background measurement and scaled to match the total measured background rate. Atmospheric neutrinos and uncertainties thereon are derived from previous measurements of both the $\pi/K$ and charm components of the atmospheric $\nu_{\mu}$ spectrum \cite{ic59_muons}. A gap larger than the one between 400 and 1000 TeV appears in 43\% of realizations of the best-fit continuous spectrum.}
\label{fig:energyhisto}
\end{figure}

\begin{figure}
\includegraphics[width=\linewidth]{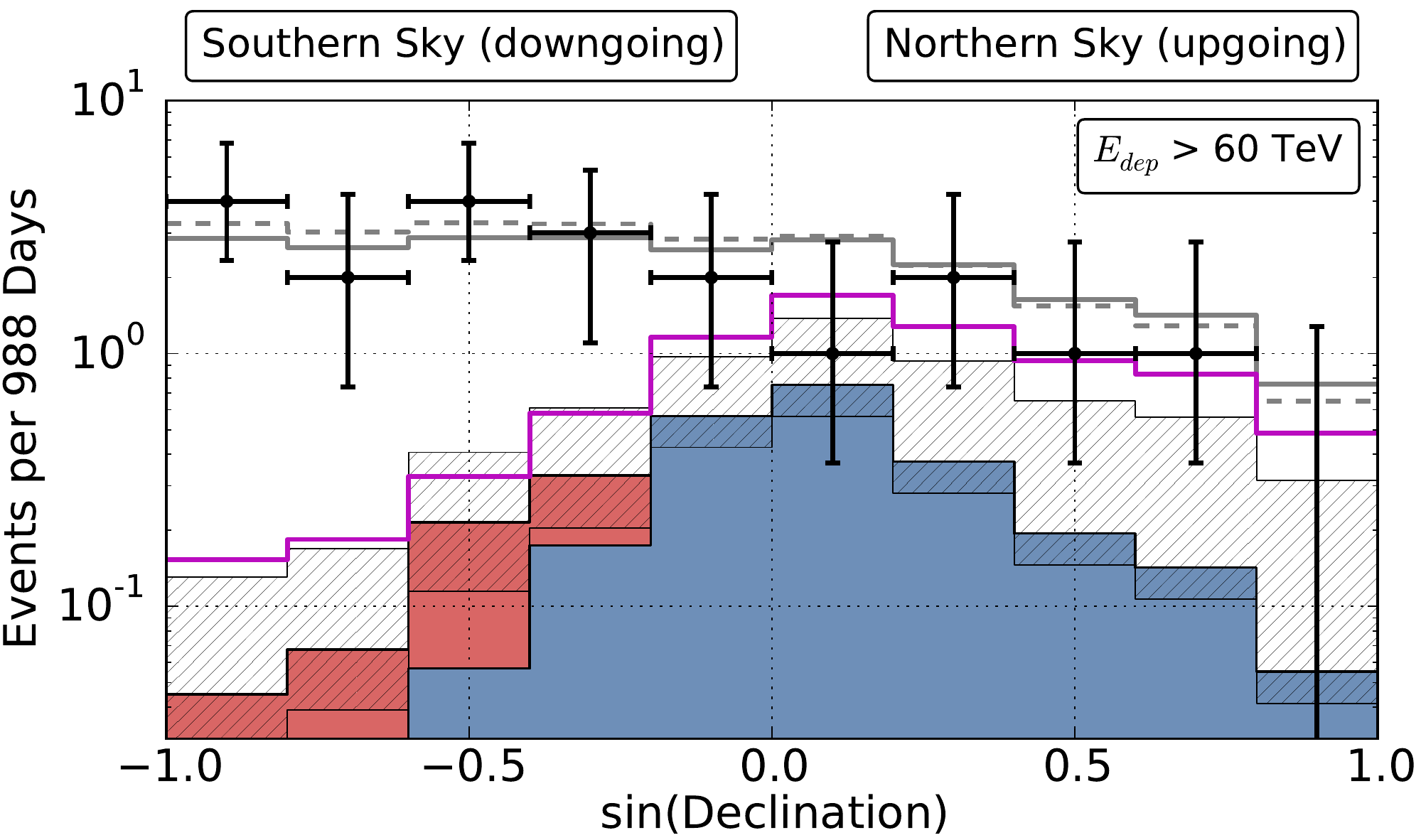}
\caption{Arrival angles of events with $E_{dep} > 60\,\mathrm{TeV}$, as used in our fit and above the majority of the cosmic ray muon background. The increasing opacity of the Earth to high energy neutrinos is visible at the right of the plot. Vetoing atmospheric neutrinos by muons from their parent air showers depresses the atmospheric neutrino background on the left. The data are described well by the expected backgrounds and a hard astrophysical isotropic neutrino flux (gray lines). Colors as in Fig.~\ref{fig:energyhisto}. Variations of this figure with other energy thresholds are in the online supplement~\cite{supplemental_section}.}
\label{fig:zenithhisto}
\end{figure}

A purely atmospheric explanation for these events is strongly disfavored by their properties. The observed deposited energy distribution extends to much higher energies (above 2 PeV, Fig.~\ref{fig:energyhisto}) than expected from the $\pi/K$ atmospheric neutrino background, which has been measured up to 100 TeV \cite{ic59_muons}. While a harder spectrum is expected from atmospheric neutrinos produced in charmed meson decay, this possibility is constrained by the observed angular distribution. Although such neutrinos are produced isotropically, approximately half \cite{atmonu_veto,newvetopaper} of those in the southern hemisphere are produced with muons of high enough energy to reach IceCube and trigger our muon veto. This results in a southern hemisphere charm rate $\sim$50\% smaller than the northern hemisphere rate, with larger ratios near the poles. Our data show no evidence of such a suppression, which is expected at some level from any atmospheric source of neutrinos (Fig.~\ref{fig:zenithhisto}).

As in \cite{hese_paper}, we quantify these arguments using a likelihood fit in arrival angle and deposited energy to a combination of background muons, atmospheric neutrinos from $\pi/K$ decay, atmospheric neutrinos from charmed meson decay, and an isotropic 1:1:1 astrophysical $E^{-2}$ test flux, as expected from charged pion decays in cosmic ray accelerators \cite{1995PhR...258..173G, 2000ARNPS..50..679L, 2002RPPh...65.1025H, 2008PhR...458..173B}. The fit included all events with $60\, \unit{TeV} < E_{dep} < 3\,\unit{PeV}$. The expected muon background in this range is below 1 event in the 3-year sample, minimizing imprecisions in modeling the muon background and threshold region. The normalizations of all background and signal neutrino fluxes were left free in the fit, without reference to uncertainties from \cite{ic59_muons}, for maximal robustness. The penetrating muon background was constrained with a Gaussian prior reflecting our veto efficiency measurement. We obtain a best-fit per-flavor astrophysical flux ($\nu + \bar \nu$) in this energy range of $E^{2} \phi(E) = 0.95 \pm 0.3 \times 10^{-8}\, \unit{GeV}\, \unit{cm}^{-2}\, \unit{s}^{-1}\, \unit{sr}^{-1}$ and background normalizations within the expected ranges. Quoted errors are $1\sigma$ uncertainties from a profile likelihood scan. This model describes the data well, with both the energy spectrum (Fig.~\ref{fig:energyhisto}) and arrival directions (Fig.~\ref{fig:zenithhisto}) of the events consistent with expectations for an origin in a hard isotropic 1:1:1 neutrino flux. The best-fit atmospheric-only alternative model, however, would require a charm normalization 3.6 times higher than our current 90\% CL upper limit from the northern hemisphere $\nu_\mu$ spectrum \cite{ic59_muons}. Even this extreme scenario is disfavored by the energy and angular distributions of the events at $5.7\sigma$ using a likelihood ratio test.

\begin{figure}
\includegraphics[width=\linewidth]{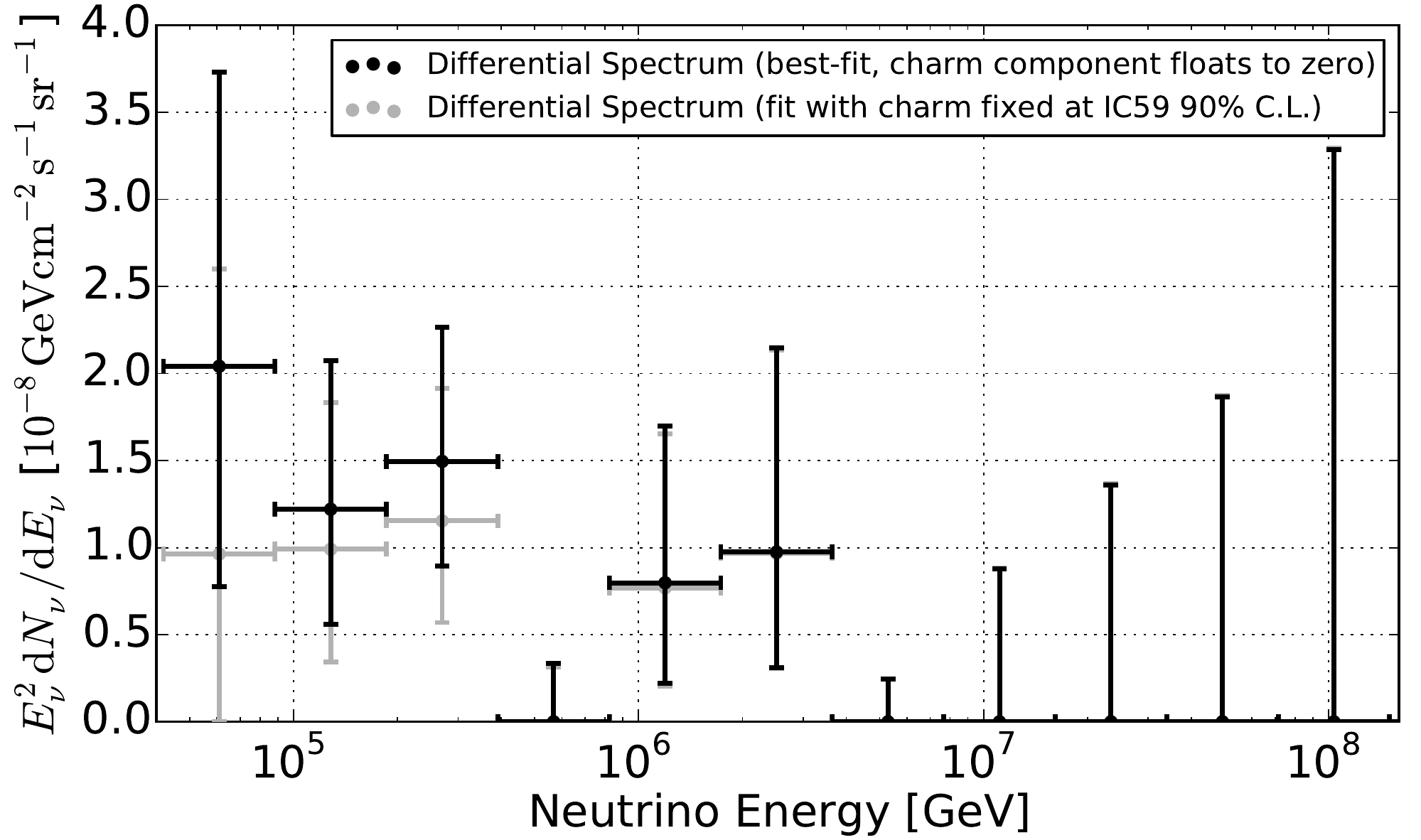}
\caption{Extraterrestrial neutrino flux ($\nu + \bar \nu$) as a function of energy. Vertical error bars indicate the $2 \Delta \mathcal L = \pm 1$ contours of the flux in each energy bin, holding all other values, including background normalizations, fixed. These provide approximate 68\% confidence ranges. An increase in the charm atmospheric background to the level of the 90\% CL limit from the northern hemisphere $\nu_\mu$ spectrum \cite{ic59_muons} would reduce the inferred astrophysical flux at low energies to the level shown for comparison in light gray. The best-fit power law is $E^2 \phi(E) = 1.5 \times 10^{-8} (E / 100 \mathrm{TeV})^{-0.3} \mathrm{GeV} \mathrm{cm}^{-2} \mathrm{s}^{-1} \mathrm{sr}^{-1}$.}
\label{fig:spectrum}
\end{figure}

Fig.~\ref{fig:spectrum} shows a fit using a more general model in which the astrophysical flux is parametrized as a piecewise function of energy rather than a continuous unbroken $E^{-2}$ power law.
As before, we assume a 1:1:1 flavor ratio and isotropy. While the reconstructed spectrum is compatible with our earlier $E^{-2}$ ansatz, an unbroken $E^{-2}$ flux at our best-fit level predicts 3.1 additional events above 2 PeV (a higher energy search \cite{ehepaper} also saw none). This may indicate, along with the slight excess in lower energy bins, either a softer spectrum or a cutoff at high energies.
Correlated systematic uncertainties in the first few points in the reconstructed spectrum (Fig.~\ref{fig:spectrum}) arise from the poorly constrained level of the charm atmospheric neutrino background. The presence of this softer ($E^{-2.7}$) component would decrease the non-atmospheric excess at low energies, hardening the spectrum of the remaining data. The corresponding range of best fit astrophysical slopes within our current 90\% confidence band on the charm flux \cite{ic59_muons} is $-2.0$ to $-2.3$. As the best-fit charm flux is zero, the best-fit astrophysical spectrum is on the lower boundary of this interval at $-2.3$ (solid line, Figs.~\ref{fig:energyhisto}, \ref{fig:zenithhisto}) with a total statistical and systematic uncertainty of $\pm 0.3$.


To identify any bright neutrino sources in the data, we employed the same maximum-likelihood clustering search as before \cite{hese_paper}, as well as searched for directional correlations with TeV gamma-ray sources.
For all tests, the test statistic (TS) is defined as the logarithm of the ratio between the best-fit likelihood including a point source component and the likelihood for the null hypothesis, an isotropic distribution \cite{ps_method}.
We determined the significance of any excess by comparing to maps scrambled in right ascension, in which our polar detector has uniform exposure. 

As in \cite{hese_paper}, the clustering analysis was run twice, first with the entire event sample, after removing the two events (28 and 32) with strong evidence of a cosmic-ray origin, and second with only the 28 shower events. This controls for bias in the likelihood fit toward the positions of single well-resolved muon tracks.
We also conducted an additional test in which we marginalize the likelihood over a uniform prior on the position of the hypothetical point source.
This reduces the bias introduced by muons, allowing track and shower events to be used together, and improves sensitivity to multiple sources by considering the entire sky rather than the single best point.

\begin{figure}
\includegraphics[width=\linewidth]{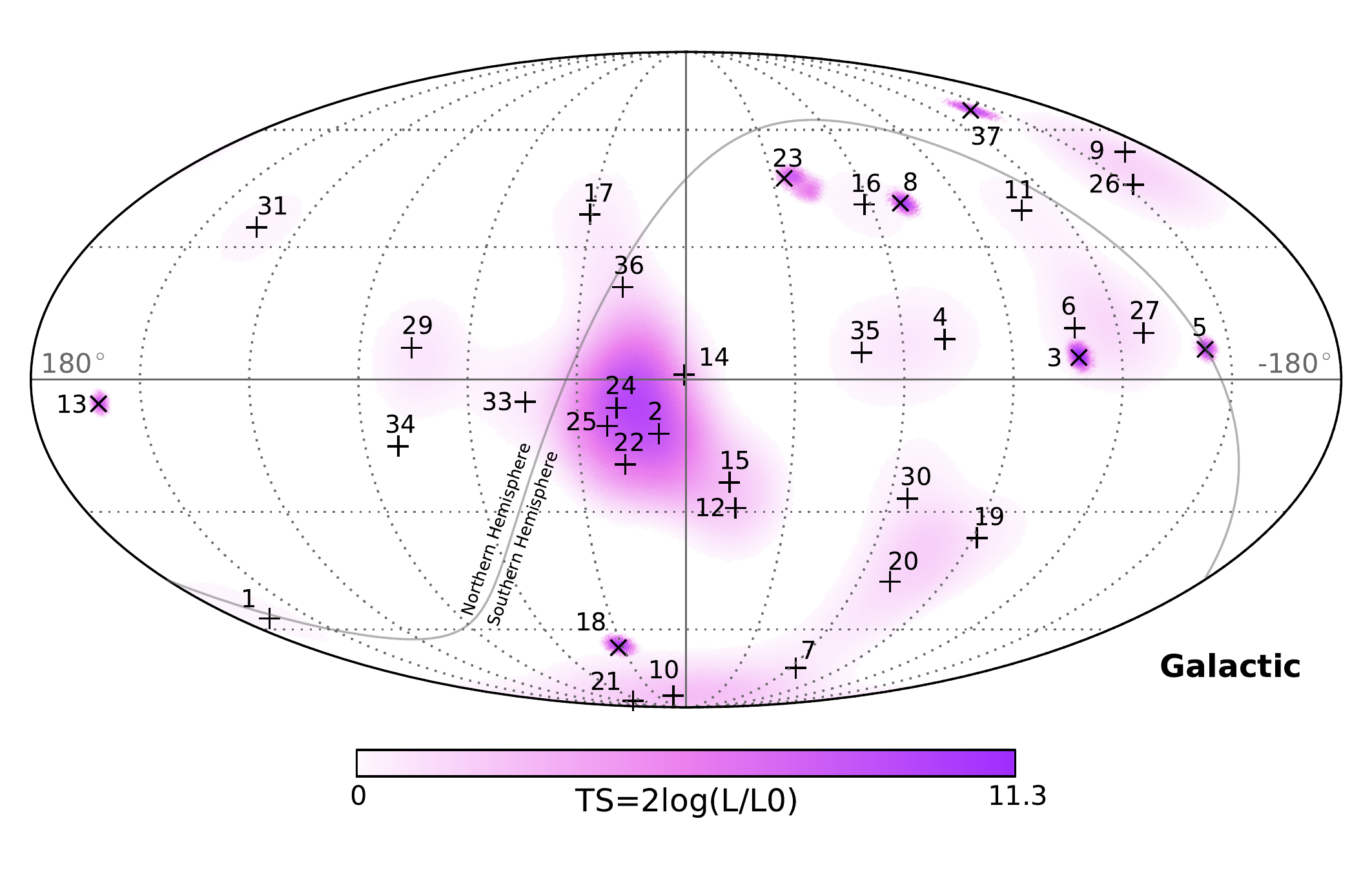}
\caption{Arrival directions of the events in galactic coordinates. Shower-like events (median angular resolution $\sim15^{\circ}$) are marked with $+$ and those containing muon tracks ($\lesssim 1^{\circ}$) with $\times$. Approximately 40\% of the events (mostly tracks \cite{2004JCAP...11..009B}) are expected to originate from atmospheric backgrounds. Event IDs match those in the catalog in the online supplement~\cite{supplemental_section} and are time ordered. The grey line denotes the equatorial plane.  Colors show the test statistic (TS) for the point source clustering test at each location. No significant clustering was observed.}
\label{fig:skymap}
\end{figure}

Three tests were performed to search for neutrinos correlated with known gamma-ray sources, also using track and shower events together.  The first two searched for clustering along the galactic plane, with a fixed width of $\pm 2.5^\circ$, based on TeV gamma-ray measurements \cite{milagroGP}, and with a free width of between $\pm 2.5^\circ$ and $\pm 30^\circ$. The last searched for correlation between neutrino events and a pre-defined catalog of potential point sources (a combination of the usual IceCube \cite{2013ApJ...779..132A} and ANTARES \cite{2012ApJ...760...53A} lists; see online supplement~\cite{supplemental_section}).  For the catalog search, the TS value was evaluated at each source location, and the post-trials significance calculated by comparing the highest observed value in each hemisphere to results from performing the analysis on scrambled datasets.

No hypothesis test yielded statistically significant evidence of clustering or correlations. 
For the all-sky clustering test (Fig.~\ref{fig:skymap}), scrambled datasets produced locations with equal or greater TS 84\% and 7.2\% of the time for all events and for shower-like events only.
As in the two-year data set, the strongest clustering was near the galactic center. Other neutrino observations of this location give no evidence for a source \cite{ANTARESPS}, however, and no new events were strongly correlated with this region.
When using the marginalized likelihood, a test statistic greater than or equal to the observed value was found in 28\% of scrambled datasets.
The source list yielded p-values for the northern and southern hemispheres of 28\% and 8\%, respectively.
Correlation with the galactic plane was also not significant: when letting the width float freely, the best fit was $\pm 7.5^{\circ}$ with a post-trials chance probability of 2.8\%, while a fixed width of $\pm 2.5^{\circ}$ returned a p-value of 24\%. A repeat of the time clustering search from \cite{hese_paper} also found no evidence for structure.


With or without a possible galactic contribution \cite{2014PhRvD..89b3501S, Ahlers:2013xia}, the high galactic latitudes of many of the highest-energy events (Fig.~\ref{fig:skymap}) suggest at least some extragalactic component. Exception may be made for local large diffuse sources (e.g. the Fermi bubbles \cite{fermi_bubbles} or the galactic halo \cite{gc_dm_halo,2014arXiv1403.3206T}), but these models typically can explain at most a fraction of the data.
If our data arise from an extragalactic flux produced by many isotropically distributed point sources, we can compare our all-sky flux with existing point-source limits.
By exploiting the additional effective volume provided by use of uncontained $\nu_\mu$ events, previous point-source studies would have been sensitive to a northern sky point source producing more than 1-10\% of our best-fit flux, depending on declination and energy spectrum \cite{pointsources}. The lack of any evidence for such sources from these studies, as well as the wide distribution of our events, thus lends support to an interpretation in terms of many individually dim sources.
Some contribution from a few comparatively bright sources cannot be ruled out, however, especially in the southern hemisphere, where the sensitivity of IceCube to point sources in uncontained $\nu_\mu$ is reduced by the large muon background and small target mass above the detector.

The neutrino spectrum (Fig.~\ref{fig:spectrum}) can also be used to constrain source properties. In almost all candidate sources \cite{SteckerAGN,2013PhRvL.111d1103K,2004NewAR..48..381A,2004PhRvD..70l3001A,1993PhRvD..47.5270N,1995APh.....3..295M,1996SSRv...75..341S,1998PhRvD..58l3005R,2001PhRvL..87q1102M,2003APh....18..593M,2005APh....23..355B,2005PhRvD..72j7301S,2012ApJ...749..155E,wb97,guetta2004,2014arXiv1401.1820B,2014arXiv1403.0574W,2000ApJ...541..707W,2003PhRvD..68h3001R,2006APh....25..118B,2006PhRvL..97e1101M,loebwaxman, 2013PhRvD..88l1301M, 2014ApJ...780..137Y,2006astro.ph..8699T}, neutrinos would be produced by the interaction of cosmic rays with either radiation or gas. Interactions with radiation ($p \gamma$) typically produce a peaked spectrum, reflecting the energy spectrum of the photons; those with gas ($pp$) produce a smooth power law \cite{Mucke:1999yb,Kelner:2006tc}. While $p \gamma$ models satisfactorily explain some aspects of the data such as the possible drop-off at high energies, many involve a central plateau smaller than our observed energy range, placing them in weak tension with the data. As an example, the $p \gamma$ AGN spectrum in \cite{SteckerAGN} peaks at several PeV with much lower predictions at 100 TeV; thus, while able to explain the highest energy events, it fits poorly at lower energies and is disfavored as the sole source at the $2 \sigma$ level with respect to our simple $E^{-2}$ test flux. Gamma-ray burst $p \gamma$ models such as \cite{wb97,guetta2004} have energy ranges better aligned with our data, with central plateaus from around 100 TeV to a few PeV, although existing limits from searches for correlations with observed GRBs are more than an order of magnitude below the observed flux \cite{ic59_grb}. Cosmic ray interactions with gas, such as predicted around supernova remnants in our and other galaxies, particularly those with high star-forming rates, produce smooth spectra with slopes reflecting post-diffusion cosmic rays (e.g. $E^{-2.2}$ in \cite{loebwaxman}) and seem to describe the data well. Large uncertainties on both the measured neutrino spectrum and all models prevent any conclusions, however.

The best-fit flux level in our central energy range ($10^{-8}\, \mathrm{GeV}\, \mathrm{cm}^{-2}\, \mathrm{s}^{-1}\, \mathrm{sr}^{-1}$ per flavor) is similar to the Waxman-Bahcall bound \cite{1999PhRvD..59b3002W}, the aggregate neutrino flux from charged pion decay in all extragalactic cosmic ray accelerators if they are optically thin. This bound is derived from the cosmic ray spectrum above $10^{18}$ eV (1000 PeV). Our neutrinos, however, are likely associated with protons at much lower energies, on the order of 1 to 10 PeV \cite{Mucke:1999yb,Kelner:2006tc}, at which the bound may be quite different \cite{2001PhRvD..63b3003M}. Along with large uncertainties in the neutrino spectrum (Fig.~\ref{fig:spectrum}), this makes correspondence with the Waxman-Bahcall bound, or $10^{18}$ eV cosmic ray sources, unclear.

Further observations with the present or upgraded IceCube detector and the planned KM3NeT \cite{KM3NeTTDR} telescope are required to answer many questions about the sources of this astrophysical flux \cite{2013PhRvD..88d3009L}.
Gamma-ray, optical, and X-ray observations of the directions of individual high-energy neutrinos, which point directly to their origins, may also be able to identify these sources even for those with neutrino luminosities too low for identification from neutrino measurements alone.

\begin{acknowledgments}
We acknowledge support from the following agencies:
U.S. National Science Foundation-Office of Polar Programs,
U.S. National Science Foundation-Physics Division,
University of Wisconsin Alumni Research Foundation,
the Grid Laboratory Of Wisconsin (GLOW) grid infrastructure at the University of Wisconsin - Madison, the Open Science Grid (OSG) grid infrastructure;
U.S. Department of Energy, and National Energy Research Scientific Computing Center,
the Louisiana Optical Network Initiative (LONI) grid computing resources;
Natural Sciences and Engineering Research Council of Canada,
WestGrid and Compute/Calcul Canada;
Swedish Research Council,
Swedish Polar Research Secretariat,
Swedish National Infrastructure for Computing (SNIC),
and Knut and Alice Wallenberg Foundation, Sweden;
German Ministry for Education and Research (BMBF),
Deutsche Forschungsgemeinschaft (DFG),
Helmholtz Alliance for Astroparticle Physics (HAP),
Research Department of Plasmas with Complex Interactions (Bochum), Germany;
Fund for Scientific Research (FNRS-FWO),
FWO Odysseus programme,
Flanders Institute to encourage scientific and technological research in industry (IWT),
Belgian Federal Science Policy Office (Belspo);
University of Oxford, United Kingdom;
Marsden Fund, New Zealand;
Australian Research Council;
Japan Society for Promotion of Science (JSPS);
the Swiss National Science Foundation (SNSF), Switzerland;
National Research Foundation of Korea (NRF);
Danish National Research Foundation, Denmark (DNRF).
Some of the results in this paper have been derived using the HEALPix \cite{healpix} package.
Thanks to R. Laha, J. Beacom, K. Murase, S. Razzaque, and N. Harrington for helpful discussions.
\end{acknowledgments}

\clearpage
\newpage
\appendix*

\ifx \standalonesupplemental\undefined
\setcounter{page}{1}
\setcounter{figure}{0}
\setcounter{table}{0}
\fi
\renewcommand{\thepage}{Supplementary Methods and Tables -- S\arabic{page}}
\renewcommand{\figurename}{SUPPL. FIG.}
\renewcommand{\tablename}{SUPPL. TABLE}


This section gives additional technical information about the result in the main article, including tabular forms of the results, alternative presentations of several figures, reviews of referenced methods, and event displays of the neutrino candidates. Some content is repeated from the main text or from our earlier publication covering the first two years of data \cite{hese_paper} for context. Methods and performance information not provided here (e.g. effective areas) are identical to those in \cite{hese_paper}. Event displays here include only the events first shown in this paper; displays for events 1-28 can be found in the online supplement to \cite{hese_paper}. Further IceCube data releases can be found at \url{http://www.icecube.wisc.edu/science/data}.

\section{Event Information}

\begin{figure}
\includegraphics[width=\linewidth]{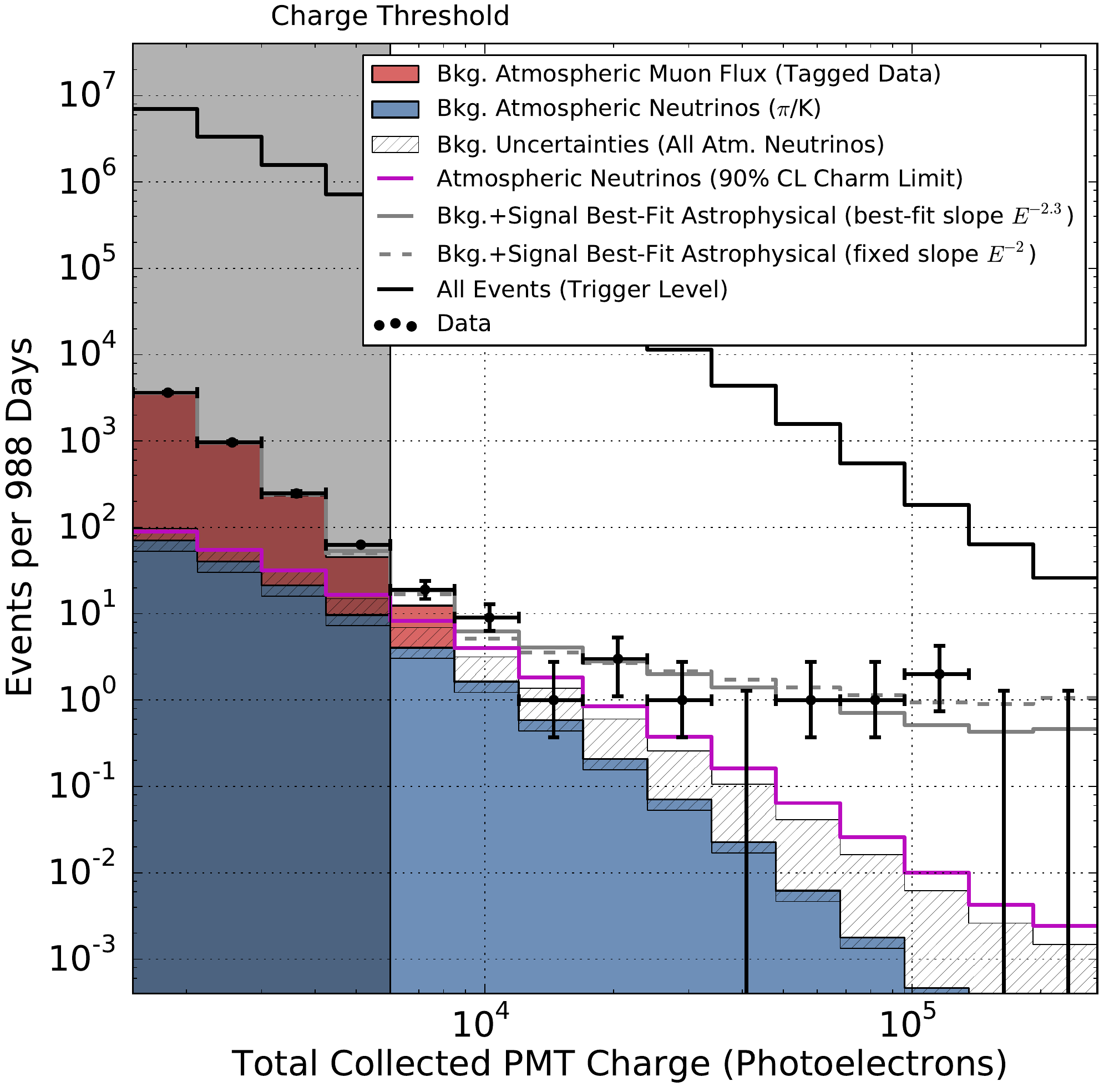}
\caption{%
Distribution of deposited PMT charges ($Q_{tot}$).
Muons at higher total charges are less likely to pass the veto layer undetected, causing the muon background (red, estimated from data) to fall faster than the overall trigger rate (uppermost line).
The data events in the unshaded region, at $Q_{tot} > 6000$, are the events reported in this work.
The hatched region shows current $1 \sigma$ experimental uncertainties on both the $\pi/K$ and prompt components of the atmospheric neutrino background \cite{ic59_muons}.
For scale, the experimental 90\% CL upper bound on prompt atmospheric neutrinos \cite{ic59_muons} is also shown (magenta line).
}
\label{fig:qtot}
\end{figure}

Properties of the 37 neutrino candidate events are shown in Suppl.~Tab.~\ref{tab:events}. Five of these (3, 8, 18, 28, 32) contain downgoing muons and have an apparent first interaction near the detector boundary and are therefore consistent with the expected $8.4 \pm 4.2$ background muon events. Two of these (28 and 32) have subthreshold early hits in the veto region, as well as coincident detections in the IceTop surface air shower array, and are almost certainly penetrating cosmic ray muon background. The remaining events are uniformly distributed throughout the detector volume and are consistent with neutrino interactions. Their distribution in total PMT charge, used for event selection, is shown in Suppl.~Fig.~\ref{fig:qtot}.

Reconstruction uncertainties given in Suppl.~Tab.~\ref{tab:events} include both statistical and systematic uncertainties and were determined from average reconstruction errors on a population of simulated events of the same topology in the same part of the detector with similar energies to those observed. The reconstructions used were maximum likelihood fits of the observed photon timing distributions to template events using the cascade and muon loss unfolding techniques described in \cite{energy_reco}. Cascade angular resolution operates by observing forward/backward asymmetries in photon timing: in front of the neutrino interaction, most light is unscattered and arrives over a short period of time, whereas behind the interaction, the light has scattered at least once, producing a broader profile. Resolution as a function of energy for this analysis is shown in Fig.~14 of \cite{energy_reco}. Event 32 is made of two coincident cosmic ray muons (see supplemental event views) and so no single energy and direction can be given for the event.

\begin{table*}
\begin{tabular}{c|c|c|c|c|c|c}
 ID  & Dep. Energy (TeV) & Observation Time (MJD)  & Decl. (deg.) & R.A. (deg.) & Med. Angular Error (deg.)  &   Event Topology \\
\hline
1 & $47.6 \,^{+6.5}_{-5.4}$ & 55351.3222143 & $-1.8$ & $35.2$ & $16.3$ & Shower \\
2 & $117 \,^{+15}_{-15}$ & 55351.4659661 & $-28.0$ & $282.6$ & $25.4$ & Shower \\
3 & $78.7 \,^{+10.8}_{-8.7}$ & 55451.0707482 & $-31.2$ & $127.9$ & $\lesssim 1.4$ & Track \\
4 & $165 \,^{+20}_{-15}$ & 55477.3930984 & $-51.2$ & $169.5$ & $7.1$ & Shower \\
5 & $71.4 \,^{+9.0}_{-9.0}$ & 55512.5516311 & $-0.4$ & $110.6$ & $\lesssim 1.2$ & Track \\
6 & $28.4 \,^{+2.7}_{-2.5}$ & 55567.6388127 & $-27.2$ & $133.9$ & $9.8$ & Shower \\
7 & $34.3 \,^{+3.5}_{-4.3}$ & 55571.2585362 & $-45.1$ & $15.6$ & $24.1$ & Shower \\
8 & $32.6 \,^{+10.3}_{-11.1}$ & 55608.8201315 & $-21.2$ & $182.4$ & $\lesssim 1.3$ & Track \\
9 & $63.2 \,^{+7.1}_{-8.0}$ & 55685.6629713 & $33.6$ & $151.3$ & $16.5$ & Shower \\
10 & $97.2 \,^{+10.4}_{-12.4}$ & 55695.2730461 & $-29.4$ & $5.0$ & $8.1$ & Shower \\
11 & $88.4 \,^{+12.5}_{-10.7}$ & 55714.5909345 & $-8.9$ & $155.3$ & $16.7$ & Shower \\
12 & $104 \,^{+13}_{-13}$ & 55739.4411232 & $-52.8$ & $296.1$ & $9.8$ & Shower \\
13 & $253 \,^{+26}_{-22}$ & 55756.1129844 & $40.3$ & $67.9$ & $\lesssim 1.2$ & Track \\
14 & $1041 \,^{+132}_{-144}$ & 55782.5161911 & $-27.9$ & $265.6$ & $13.2$ & Shower \\
15 & $57.5 \,^{+8.3}_{-7.8}$ & 55783.1854223 & $-49.7$ & $287.3$ & $19.7$ & Shower \\
16 & $30.6 \,^{+3.6}_{-3.5}$ & 55798.6271285 & $-22.6$ & $192.1$ & $19.4$ & Shower \\
17 & $200 \,^{+27}_{-27}$ & 55800.3755483 & $14.5$ & $247.4$ & $11.6$ & Shower \\
18 & $31.5 \,^{+4.6}_{-3.3}$ & 55923.5318204 & $-24.8$ & $345.6$ & $\lesssim 1.3$ & Track \\
19 & $71.5 \,^{+7.0}_{-7.2}$ & 55925.7958619 & $-59.7$ & $76.9$ & $9.7$ & Shower \\
20 & $1141 \,^{+143}_{-133}$ & 55929.3986279 & $-67.2$ & $38.3$ & $10.7$ & Shower \\
21 & $30.2 \,^{+3.5}_{-3.3}$ & 55936.5416484 & $-24.0$ & $9.0$ & $20.9$ & Shower \\
22 & $220 \,^{+21}_{-24}$ & 55941.9757813 & $-22.1$ & $293.7$ & $12.1$ & Shower \\
23 & $82.2 \,^{+8.6}_{-8.4}$ & 55949.5693228 & $-13.2$ & $208.7$ & $\lesssim 1.9$ & Track \\
24 & $30.5 \,^{+3.2}_{-2.6}$ & 55950.8474912 & $-15.1$ & $282.2$ & $15.5$ & Shower \\
25 & $33.5 \,^{+4.9}_{-5.0}$ & 55966.7422488 & $-14.5$ & $286.0$ & $46.3$ & Shower \\
26 & $210 \,^{+29}_{-26}$ & 55979.2551750 & $22.7$ & $143.4$ & $11.8$ & Shower \\
27 & $60.2 \,^{+5.6}_{-5.6}$ & 56008.6845644 & $-12.6$ & $121.7$ & $6.6$ & Shower \\
28 & $46.1 \,^{+5.7}_{-4.4}$ & 56048.5704209 & $-71.5$ & $164.8$ & $\lesssim 1.3$ & Track \\
29 & $32.7 \,^{+3.2}_{-2.9}$ & 56108.2572046 & $41.0$ & $298.1$ & $7.4$ & Shower \\
30 & $129 \,^{+14}_{-12}$ & 56115.7283574 & $-82.7$ & $103.2$ & $8.0$ & Shower \\
31 & $42.5 \,^{+5.4}_{-5.7}$ & 56176.3914143 & $78.3$ & $146.1$ & $26.0$ & Shower \\
32 & --- & 56211.7401231 & --- & --- & --- & Coincident \\
33 & $385 \,^{+46}_{-49}$ & 56221.3424023 & $7.8$ & $292.5$ & $13.5$ & Shower \\
34 & $42.1 \,^{+6.5}_{-6.3}$ & 56228.6055226 & $31.3$ & $323.4$ & $42.7$ & Shower \\
35 & $2004 \,^{+236}_{-262}$ & 56265.1338677 & $-55.8$ & $208.4$ & $15.9$ & Shower \\
36 & $28.9 \,^{+3.0}_{-2.6}$ & 56308.1642740 & $-3.0$ & $257.7$ & $11.7$ & Shower \\
37 & $30.8 \,^{+3.3}_{-3.5}$ & 56390.1887627 & $20.7$ & $167.3$ & $\lesssim 1.2$ & Track \\
\end{tabular}
\caption{Properties of the events. Tabular form of Fig.~\ref{fig:energyzenith}. Events 1-28 were previously published in \cite{hese_paper} and are included here, with no changes, for completeness. Events 28 and 32 have coincident hits in the IceTop surface array, implying that they are almost certainly produced in cosmic ray air showers.}
\label{tab:events}
\end{table*}

\section{Point Source Methods}

The point source searches used the unbinned maximum likelihood method from \cite{ps_method}:

\begin{equation}
\mathcal{L}(n_s,\vec{x_s}) = \prod_{i=1}^{N} \Big[\frac{n_s}{N}\mathcal{S}_i(\vec{x_s}) + (1 - \frac{n_s}{N})\mathcal{B}_i\Big].
\label{eq:lh}
\end{equation}

\begin{figure}
\includegraphics[width=\linewidth]{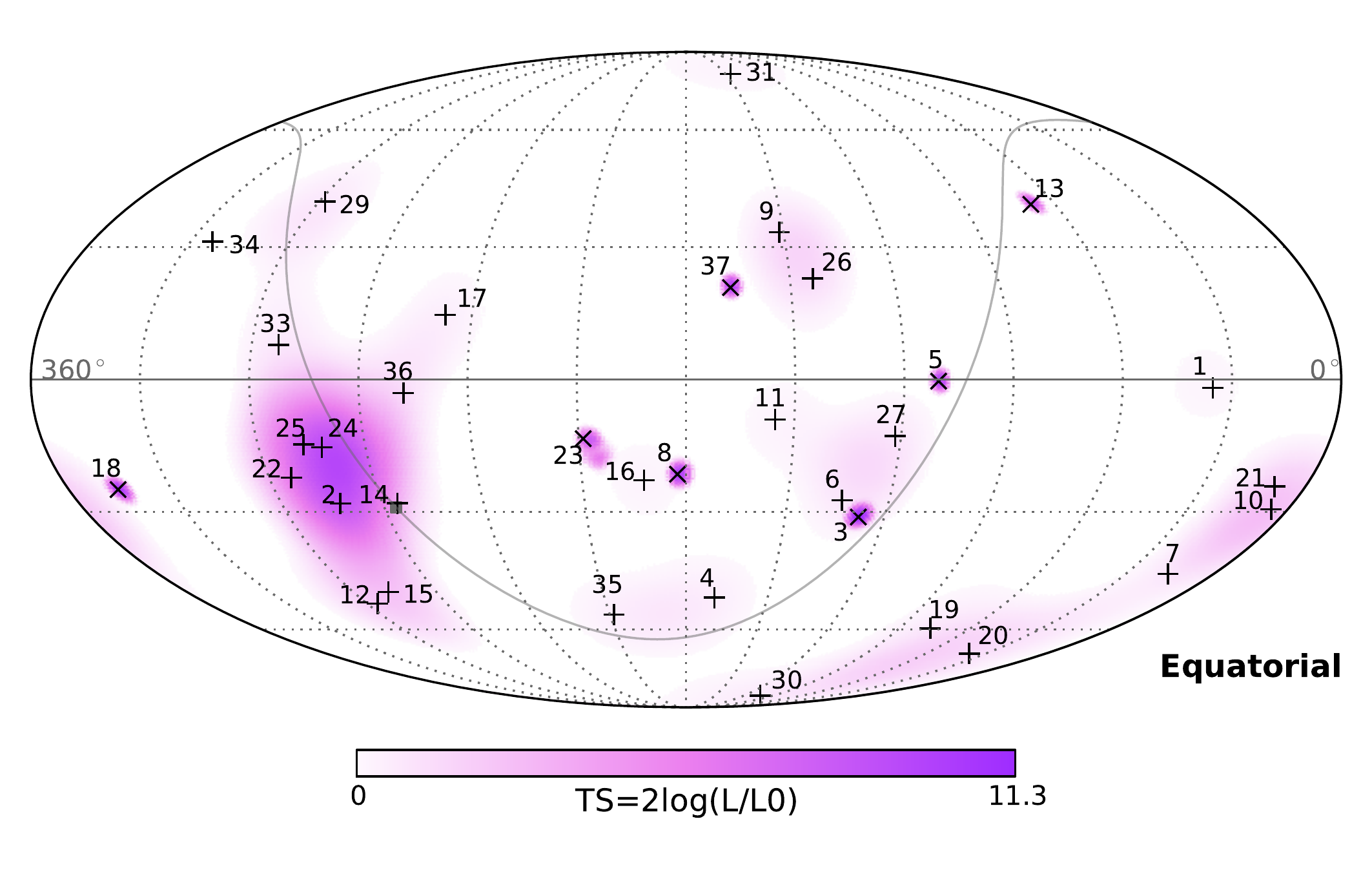}
\caption{Arrival directions of the events ($+$ for shower events, $\times$ for track events) and test statistic (colors) in equatorial coordinates (J2000).  The gray line denotes the galactic plane. This is an equatorial version of Fig.~\ref{fig:skymap}.}
\label{fig:skymap_equatorial}
\end{figure}

Here, $\mathcal{B}_i=\frac{1}{4\pi}$ represents the isotropic background probability distribution function (PDF), and the signal PDF $\mathcal{S}_i$ is the reconstructed directional uncertainty map for each event.  $N$ is the total number of events in the data sample and $n_s$ is the number of signal events, which is a free parameter.  For the all-sky clustering and source catalog searches, the likelihood is maximized at each location, resulting in a best-fit \# of signal events.  Suppl.~Fig.~\ref{fig:skymap_equatorial} shows the arrival directions of the events and the result of the point source clustering test in equatorial coordinates (J2000), while Suppl.~Tab.~\ref{tab:source_list_north} and ~\ref{tab:source_list_south} list the results for the 78 sources in the pre-defined catalog.  This catalog was chosen based on gamma-ray observations or predicted astrophysical neutrino fluxes, and is comprised of sources previously tested by IceCube \cite{2013ApJ...779..132A} and ANTARES \cite{2012ApJ...760...53A}.

To reduce the bias in the likelihood fit towards positions of single well-resolved muon tracks, a marginalized form of the likelihood was also used for the all-sky test:

\begin{equation}
\mathcal{L}_{M}(n_s) = \int\limits_{\vec{x_s}} \mathcal{L}(n_s,\vec{x_s})P(\vec{x_s})d\vec{x_s},
\label{marginalllh}
\end{equation}

where $\mathcal{L}(n_s,\vec{x_s})$ is equation \ref{eq:lh} and $P(\vec{x_s})=\frac{1}{4\pi}$ is a uniform prior for the position of a single point source.  In this procedure, there is only one free parameter ($n_s$) that is fit across the entire sky, instead of being varied independently at every position.

\begin{table}
\begin{tabular}{| l | l | c | c | c | c |}
\hline Category & Source & RA ($^{\circ}$) & Dec ($^{\circ}$) & ${\hat n}_s$ & p-value\\
\hline
SNR    &             TYCHO &   6.36 & 64.18 &  0.0 & --\\
&             Cas A & 350.85 & 58.82 &  0.0 & --\\
&             IC443 &  94.18 & 22.53 &  0.0 & --\\
&         W51C & 290.75 & 14.19 &  0.7 & 0.05\\
&               W44 & 284.04 &  1.38 &  2.5 & 0.01\\
&                W28 & 270.43 & -23.34 &  4.3 & 0.01\\
&    RX J1713.7-3946 & 258.25 & -39.75 &  0.0 & --\\
&    RX J0852.0-4622 & 133.0  & -46.37 &  0.0 & --\\
&             RCW 86 & 220.68 & -62.48 &  0.3 & 0.41\\
\hline
XB/mqso    &          LSI 303 &  40.13 & 61.23 &  0.0 & --\\
   &   Cyg X-3 & 308.10 & 41.23 &  0.8 & 0.05\\\
&           Cyg X-1 & 299.59 & 35.20 &  1.0 & 0.03\\
&    HESS J0632+057 &  98.24 &  5.81 &  0.0 & --\\
&             SS433 & 287.96 &  4.98 &  1.5 & 0.02\\
&            LS 5039 & 276.56 & -14.83 & 4.9 & 0.002\\
&           GX 339-4 & 255.7  & -48.79 &  0.0 & --\\
&            Cir X-1 & 230.17 & -57.17 &  0.0 & --\\
   \hline
Star Form-  &   Cyg OB2 & 308.10 & 41.23 &  0.8 & 0.05\\
 ation Region & & & & & \\
 \hline
Pulsar/PWN   &     MGRO J2019+37 & 305.22 & 36.83 &  0.9 & 0.04\\
&       Crab Nebula &  83.63 & 22.01 &  0.0 & --\\
&           Geminga &  98.48 & 17.77 &  0.0 & --\\
&    HESS J1912+101 & 288.21 & 10.15 &  0.8 & 0.04\\
&             Vela X & 128.75 & -45.6  &  0.0 & --\\
&     HESS J1632-478 & 248.04 & -47.82 &  0.0 & --\\
&     HESS J1616-508 & 243.78 & -51.40 &  0.0 & --\\
&     HESS J1023-575 & 155.83 & -57.76 &  0.2 & 0.44\\
&          MSH 15-52 & 228.53 & -59.16 &  0.06 & 0.48\\
&     HESS J1303-631 & 195.74 & -63.52 &  0.8 & 0.28\\
&     PSR B1259-63 & 195.74 & -63.52 &  0.8 & 0.28\\
&     HESS J1356-645 & 209.0  & -64.5  &  0.5 & 0.35\\
   \hline
Galactic  &             Sgr A* & 266.42 & -29.01 &  3.1 & 0.04\\
Center  &    & &  &  &\\
\hline
Not  &     MGRO J1908+06 & 286.99 &  6.27 &  1.3 & 0.03\\
Identified &     HESS J1834-087 & 278.69 &  -8.76 &  4.7 & 0.01\\
&     HESS J1741-302 & 265.25 & -30.2  &  2.5 & 0.07\\
&     HESS J1503-582 & 226.46 & -58.74 &  0.2 & 0.45\\
&     HESS J1507-622 & 226.72 & -62.34 &  0.1 & 0.47\\

\hline
\end{tabular}

\caption{Catalog of 36 galactic sources, grouped according to their classification as supernova remnants (SNR), X-ray binaries or microquasars (XB/mqso), pulsar wind nebulae (PWN), star formation regions, and unidentified sources.  The post-trials p-values for the entire catalog in the northern and southern hemispheres were 28\% and 8\%, respectively.  For each source, the pre-trials p-value was estimated by repeating the source catalog search with the data randomized in right ascension.  The fraction of
test statistic (TS) values from all individual sources that were greater than or equal to the observed TS determined the pre-trials p-value.  The best-fit \# of signal events ($\hat{n}_s$) is the result of the likelihood fit at each individual source.  When $\hat{n}_s$ = 0, no p-value is reported.  Since many sources are spatially close together relative to the angular resolution, adjacent sources often receive similar fit results.  For sources separated by less than $1^{\circ}$, their positions are averaged and they are treated as one source.}
\label{tab:source_list_north}
\end{table}

\begin{table}
\begin{tabular}{| l | l | c | c | c | c |}
\hline Category & Source & RA ($^{\circ}$) & Dec ($^{\circ}$) & $\hat{n}_s$ & p-value\\
\hline
BL Lac & S5 0716+71 & 110.47 & 71.34 &  0.0 & --\\
&      1ES 1959+650 & 300.00 & 65.15 &  0.0 & --\\
&      1ES 2344+514 & 356.77 & 51.70 &  0.0 & --\\
&             3C66A &  35.67 & 43.04 &  0.0 & --\\
&       H 1426+428 & 217.14 & 42.67 &  0.0 & --\\
&            BL Lac & 330.68 & 42.28 &  0.0 & --\\
&           Mrk 501 & 253.47 & 39.76 &  0.0 & --\\
&           Mrk 421 & 166.11 & 38.21 &  0.0 & --\\
&           W Comae & 185.38 & 28.23 &  0.0 & --\\
&      1ES 0229+200 &  38.20 & 20.29 &  0.0 & --\\
&     PKS 0235+164 &  39.66 & 16.62 &  0.0 & --\\
&     VER J0648+152 & 102.2  & 15.27 &  0.0 & --\\
&     RGB J0152+017 &  28.17 &  1.79 &  0.1 & 0.15\\
&       1ES 0347-121 &  57.35 & -11.99 &  0.0 & --\\
&       1ES 1101-232 & 165.91 & -23.49 &  0.0 & --\\
&       PKS 2155-304 & 329.72 & -30.22 &  0.0 & --\\
&         H 2356-309 & 359.78 & -30.63 &  1.8 & 0.08\\
&       PKS 0548-322 &  87.67 & -32.27 &  0.0 & --\\
&       PKS 0426-380 &  67.17 & -37.93 &  0.0 & --\\
&       PKS 0537-441 &  84.71 & -44.08 &  0.0 & --\\
&       PKS 2005-489 & 302.37 & -48.82 &  1.5 & 0.11\\
\hline
FSRQ   &          4C 38.41 & 248.82 & 38.14 &  0.0 & --\\
&          3C 454.3 & 343.50 & 16.15 &  0.0 & --\\
&      PKS 0528+134 &  82.74 & 13.53 &  0.0 & --\\
&      PKS 1502+106 & 226.10 & 10.52 &  0.0 & --\\
&            3C 273 & 187.28 &  2.05 &  0.0 & --\\
&              3C279 & 194.05 &  -5.79 &  0.0 & --\\
&     HESS J1837-069 & 279.41 &  -6.95 &  4.5 & 0.01\\
&       QSO 2022-077 & 306.42 &  -7.64 &  0.4 & 0.44\\
&       PKS 1406-076 & 212.24 &  -7.87 &  0.0 & --\\
&       PKS 0727-11 & 112.58 & -11.7  &  0.4 & 0.39\\
&       QSO 1730-130 & 263.26 & -13.08 &  3.3 & 0.03\\
&       PKS 0454-234 &  74.27 & -23.43 &  0.0 & --\\
&       PKS 1622-297 & 246.53 & -29.86 &  0.0 & --\\
&       PKS 1454-354 & 224.36 & -35.65 &  0.0 & --\\
\hline
Starburst   &               M82 & 148.97 & 69.68 &  0.07 & 0.15\\
\hline
Radio &  NGC 1275 &  49.95 & 41.51 &  0.0 & --\\
Galaxies &   Cyg A & 299.87 & 40.73 &  0.9 & 0.03\\
&          3C 123.0 &  69.27 & 29.67 &  0.0 & --\\
&               M87 & 187.71 & 12.39 &  0.0 & --\\
&              Cen A & 201.37 & -43.02 &  0.03 & 0.49\\
\hline
Seyfert & ESO 139-G12 & 264.41 & -59.94 &  0.0 & --\\
\hline
\end{tabular}
\caption{Catalog of 42 extragalactic sources, grouped according to their classification as BL Lac objects, Radio galaxies, Flat Spectrum Radio Quasars (FSRQ), Starburst galaxies, and Seyfert galaxies.  A description of the information in the table can be found in Suppl.~Tab.~\ref{tab:source_list_north}.}
\label{tab:source_list_south}
\end{table}

For the galactic plane search, equation \ref{eq:lh} is modified so the signal PDF only includes regions which overlap with the galactic plane:

\begin{equation}
\mathcal{S}_i(\vec{x_j}) \rightarrow \sum_j^{n_{bins}}\frac{W(\vec{x_j})\mathcal{S}_i(\vec{x_j})}{n_{bins}}.
\label{galplane}
\end{equation}

\begin{figure}
\includegraphics[width=\linewidth]{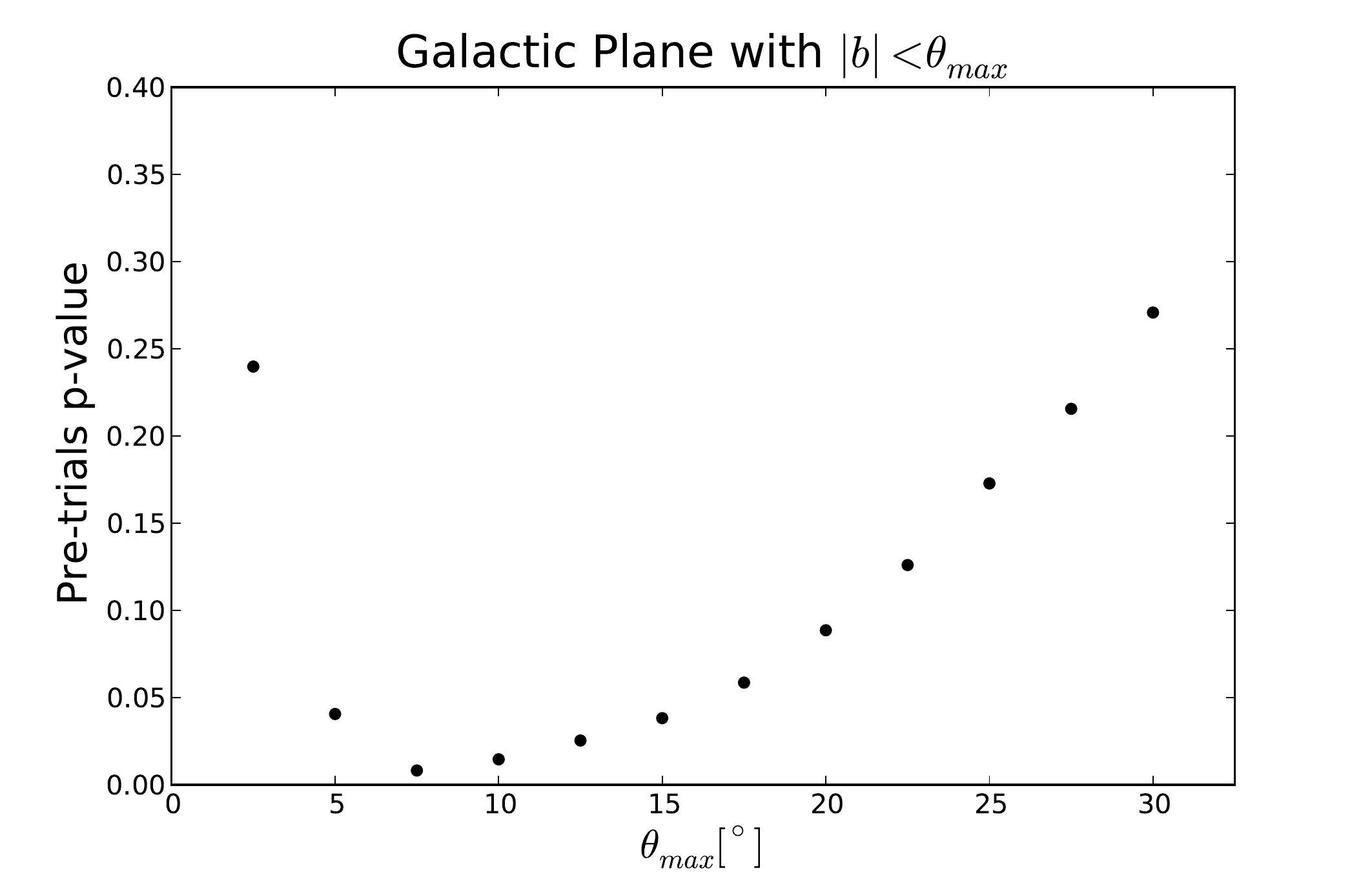}
\caption{Pre-trials p-value vs. width of galactic plane hypothesis.  The width of the galactic plane is varied from $\pm 2.5^{\circ}$ to $\pm 30^{\circ}$ in steps of $2.5^{\circ}$.  For each width, the pre-trials p-value is calculated by comparing the maximized likelihood to that from scrambled datasets.  All results are consistent with the background-only hypothesis.}
\label{fig:galactic_plane_scan}
\end{figure}

The weight $W(\vec{x_j})$ is set to 1 for any region overlapping a galactic plane with a specific angular extent, and is set to 0 otherwise.  Suppl.~Fig.~\ref{fig:galactic_plane_scan} shows the degree of clustering along the galactic plane for each tested width of the plane.

\section{Alternative Hypothesis Tests}

The primary statistical test used in this article is based on optimization of a Poisson likelihood in zenith angle and deposited energy containing four components: penetrating muon background, atmospheric neutrinos from $\pi/K$ decay, atmospheric neutrinos from charm decay, and an isotropic $E^{-2}$ astrophysical test flux. The muon background was constrained by a Gaussian prior matching our veto efficiency measurement. To ensure maximum robustness, all neutrino rates were completely unconstrained beyond a non-negativity requirement.

\begin{figure}
\includegraphics[width=\linewidth]{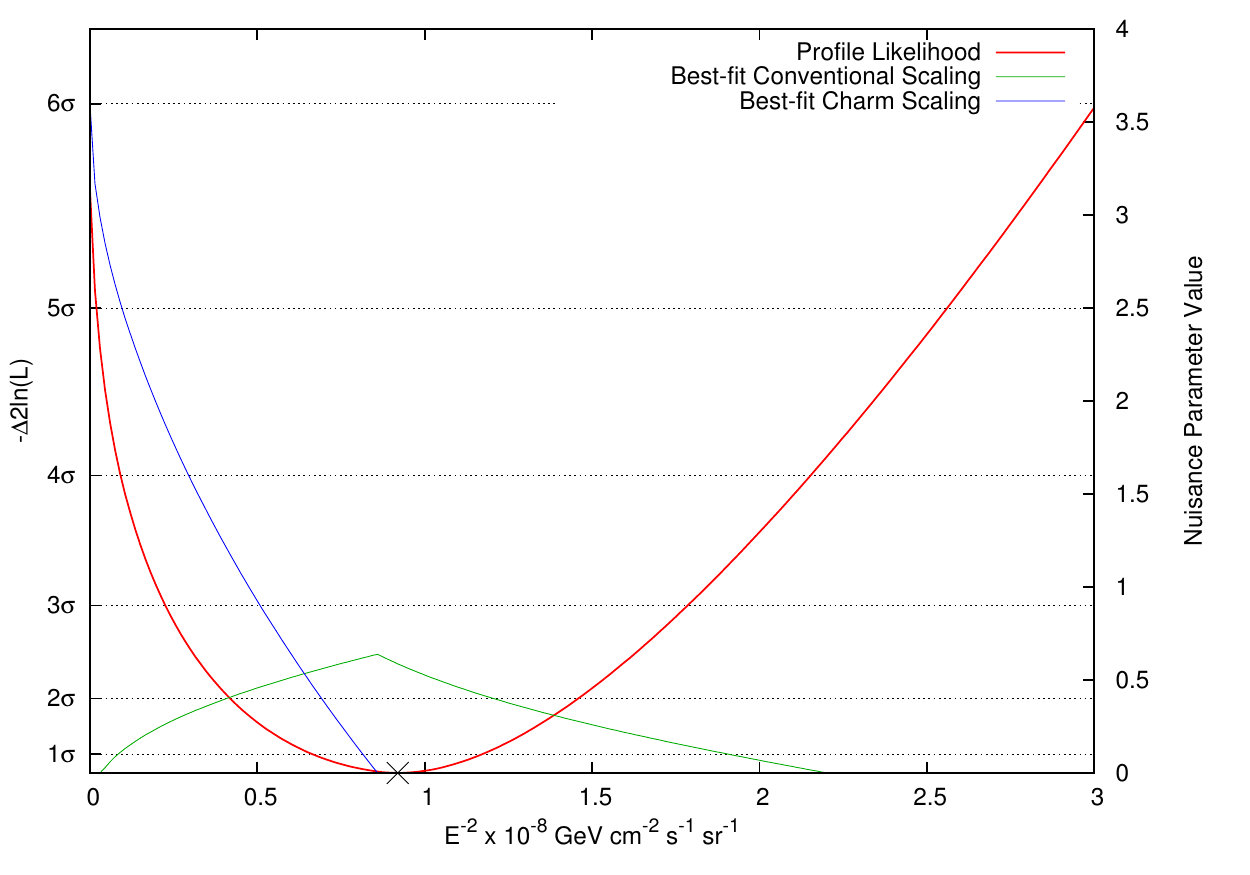}
\caption{Profile likelihood scan of the normalization of the $E^{-2}$ test flux for the unconstrained fit. The red line represents the likelihood difference (left axis) to the best-fit point (marked with $\times$). Nuisance parameters (right axis, blue and green lines) are fractions of, respectively, the 90\% CL upper limit on prompt and best-fit conventional ($\pi/K$) atmospheric neutrino fluxes from \cite{ic59_muons} and show the best-fit values, without uncertainties, of the atmospheric flux for each choice of astrophysical flux. For very low astrophysical fluxes, large prompt atmospheric neutrino fluxes are required to explain the data (blue line) but even large values are in strong tension with the data (red line). Note that significances given on the left axis are approximate, although they coincide with results of Monte Carlo ensembles for the null hypothesis rejection ($5.7\sigma$).}
\label{fig:llhspace}
\end{figure}

To test the null hypothesis of no astrophysical flux, we compared the best global fit, with all components free, to the best fit when the astrophysical test flux was constrained to zero using the difference in likelihood as a test statistic. This rejected with a significance of $5.7 \sigma$ the no-astrophysical case when compared to the best-fit alternative, which had a prompt flux (the hardest non-astrophysical component available to the fitter) 3.6 times above existing 90\% CL limits \cite{ic59_muons} (Suppl.~Fig.~\ref{fig:llhspace}), which themselves are well above most common prompt flux predictions (e.g. \cite{2008PhRvD..78d3005E}). Using the previous limits directly in the fit, through a Gaussian penalty function, would have increased the significance of the result to $6.8 \sigma$, tested against a best-fit prompt flux 1.6 times larger than the existing 90\% CL limit.

In the first part of this study \cite{hese_paper}, we performed an additional test that does not include information on the spectrum or angular distribution of the penetrating muon background and has correspondingly much lower sensitivity. The construction of the test also does not allow incorporation of any non-statistical uncertainties in the atmospheric neutrino fluxes, in order to match the treatment and charm background model in \cite{ehepaper}; it is presented here only for consistency with the previous result. Removing the two $\sim 1$ PeV events from the sample and incorporating them with the significance from \cite{ehepaper} gives $4.8 \sigma$. Including all events directly in the test yields $5.2 \sigma$.

\begin{figure*}
\includegraphics[width=\linewidth]{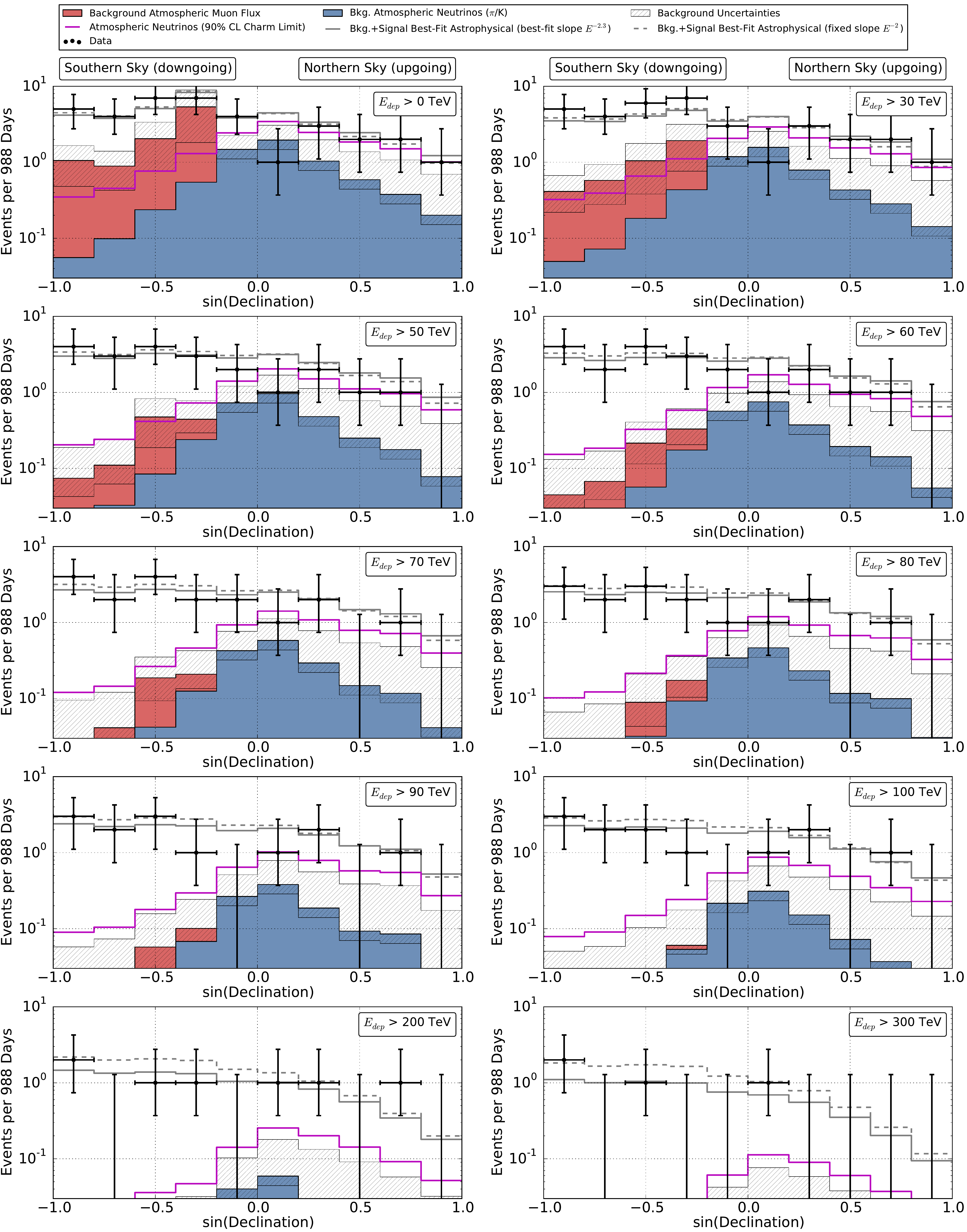}
\caption{Expected and observed distribution of events in declination for various cuts in deposited energy. The solid gray line ($E^{-2.3}$ added to backgrounds) provides a better fit to the data than the $E^{-2}$ benchmark (dashed) at the $1 \sigma$ level.}
\label{fig:zenithdistslices}
\end{figure*}

\begin{figure}
\includegraphics[width=\linewidth]{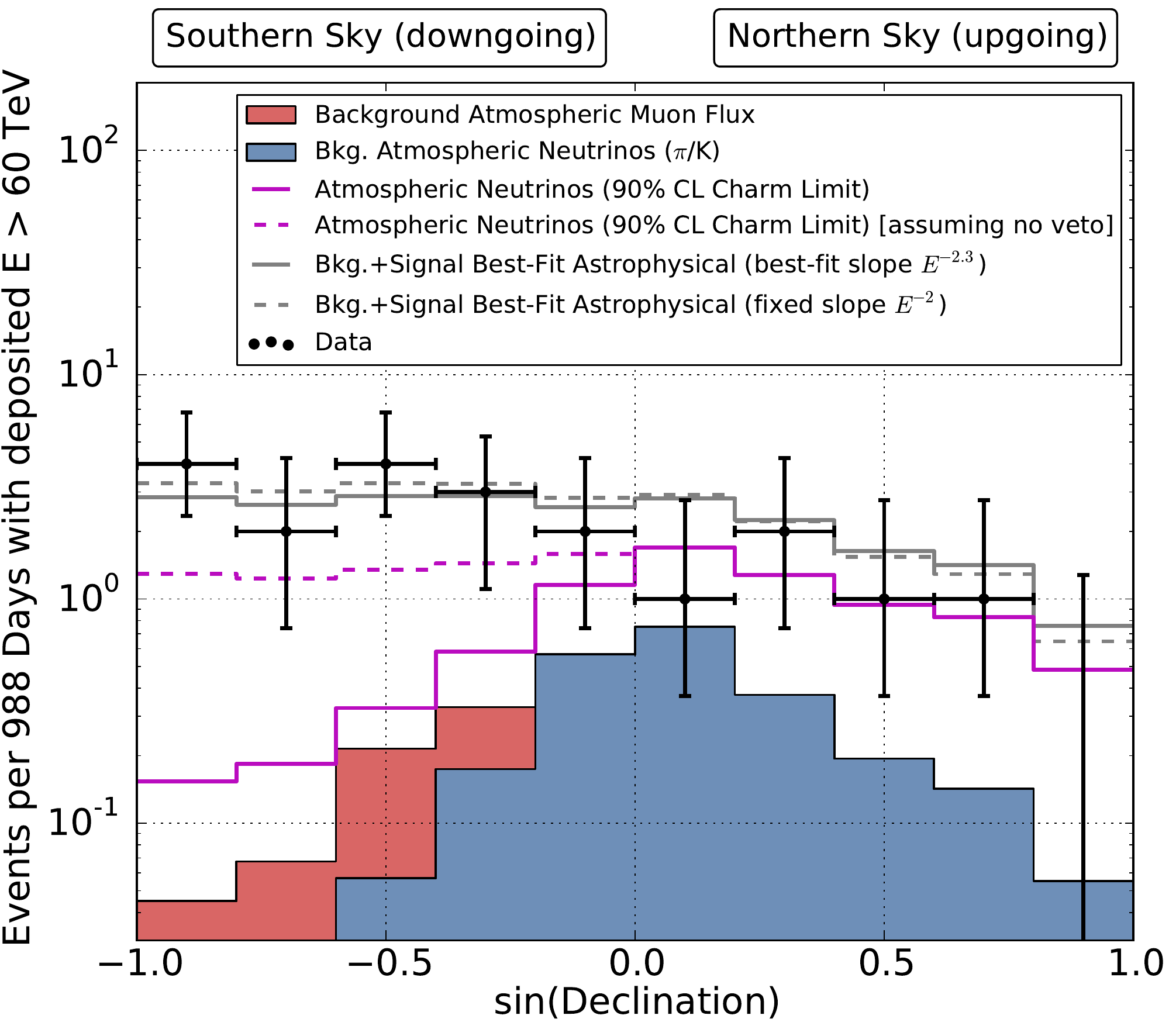}
\caption{Comparison of zenith distributions for atmospheric neutrino flux with charm saturating previous limits \cite{ic59_muons} before (dashed purple line) and after (solid purple line) removal of events accompanied into the detector by muons from the neutrinos' parent air shower \cite{atmonu_veto,newvetopaper}.}
\label{fig:charmwithoutveto}
\end{figure}

Comparisons of the properties of the events to model expectations are given in Suppl.~Tab.~\ref{tab:expectations} and Suppl.~Figs. \ref{fig:zenithdistslices} and \ref{fig:charmwithoutveto}.

\begin{table*}
\begin{tabular}{c|c|c|c|c|c|c|c|c}
\multicolumn{9}{c}{ \textbf{all energies} } \\
                             & Muons         & $\pi/K$ atm. $\nu$  & Prompt atm. $\nu$ & $E^{-2}$ (best-fit) & $E^{-2.3}$ (best-fit) & Sum ($E^{-2}$) & Sum ($E^{-2.3}$) & Data \\
\hline
Tot. Events                  & $8.4 \pm 4.2$ & $6.6_{-1.6}^{+2.2}$ & $< 9.0$ (90\% CL) & 23.8                & 23.7                  & 38.8           & 38.7             & 37 (36)   \\
Up                           & 0             & 4.2                 & $< 6.1$           & 8.3                 & 9.4                   & 12.4           & 13.5             & 9    \\
Down                         & 8.4           & 2.4                 & $< 2.9$           & 15.5                & 14.4                  & 26.3           & 25.2             & 27   \\
Track                        & $\sim 7.6$    & 4.5                 & $< 1.7$           & 4.6                 & 4.3                   & 16.7           & 16.4             & 8    \\
Shower                       & $\sim 0.8$    & 2.1                 & $< 7.2$           & 19.2                & 19.5                  & 22.1           & 22.4             & 28   \\
\hline
Fraction Up                  & 0\%           & 63\%                & 68\%              & 35\%                & 40\%                  & 32\%           & 35\%             & 25\% \\
Fraction Down                & 100\%         & 37\%                & 32\%              & 65\%                & 60\%                  & 68\%           & 65\%             & 75\% \\
Fraction Tracks              & $ > 90 \%$    & 69\%                & 19\%              & 19\%                & 18\%                  & 43\%           & 42\%             & 24\% \\
Fraction Showers             & $ < 10 \%$    & 31\%                & 81\%              & 81\%                & 82\%                  & 57\%           & 58\%             & 76\% \\

\multicolumn{9}{c}{ \; } \\
\multicolumn{9}{c}{ $\mathbf{ E_{dep} < 60 }$ \textbf{TeV} } \\
                             & Muons         & $\pi/K$ atm. $\nu$  & Prompt atm. $\nu$ & $E^{-2}$ (best-fit) & $E^{-2.3}$ (best-fit) & Sum ($E^{-2}$) & Sum ($E^{-2.3}$) & Data \\
\hline
Tot. Events                  & 8.0           & 4.2                 & $< 3.7$           & 2.2                 & 3.8                   & 14.5           & 16.1             & 16   \\
Up                           & 0             & 2.6                 & $< 2.4$           & 1.2                 & 2.0                   & 3.7            & 4.7              & 4    \\
Down                         & 8.0           & 1.6                 & $< 1.3$           & 1.1                 & 1.8                   & 10.7           & 11.4             & 12   \\
Track                        & $\sim 7.2$    & 2.9                 & $< 0.7$           & 0.4                 & 0.6                   & 10.5           & 10.7             & 4    \\
Shower                       & $\sim 0.8$    & 1.4                 & $< 3.0$           & 1.8                 & 3.2                   & 4.0            & 5.3              & 12   \\
\hline
Fraction Up                  & 0\%           & 63\%                & 65\%              & 52\%                & 53\%                  & 26\%           & 29\%             & 25\% \\
Fraction Down                & 100\%         & 37\%                & 35\%              & 48\%                & 47\%                  & 74\%           & 71\%             & 75\% \\
Fraction Tracks              & $ > 90 \%$    & 68\%                & 19\%              & 19\%                & 17\%                  & 72\%           & 67\%             & 25\% \\
Fraction Showers             & $ < 10 \%$    & 32\%                & 81\%              & 81\%                & 83\%                  & 28\%           & 33\%             & 75\% \\

\multicolumn{9}{c}{ \; } \\
\multicolumn{9}{c}{ \textbf{60 TeV} $\mathbf{ < E_{dep} < }$ \textbf{3 PeV} } \\
                             & Muons         & $\pi/K$ atm. $\nu$  & Prompt atm. $\nu$ & $E^{-2}$ (best-fit) & $E^{-2.3}$ (best-fit) & Sum ($E^{-2}$) & Sum ($E^{-2.3}$) & Data \\
\hline
Tot. Events                  & 0.4           & 2.4                 & $< 5.3$           & 18.2                & 18.6                  & 21.0           & 21.4             & 20   \\
Up                           & 0             & 1.5                 & $< 3.7$           & 6.7                 & 7.2                   & 8.2            & 8.7              & 5    \\
Down                         & 0.4           & 0.8                 & $< 1.6$           & 11.6                & 11.4                  & 12.8           & 12.7             & 15   \\
Track                        & $\sim 0.4$    & 1.7                 & $< 1.0$           & 3.8                 & 3.5                   & 5.8            & 5.5              & 4    \\
Shower                       & $\sim 0.0$    & 0.7                 & $< 4.2$           & 14.4                & 15.1                  & 15.2           & 15.8             & 16   \\
\hline
Fraction Up                  & 0\%           & 64\%                & 70\%              & 37\%                & 39\%                  & 39\%           & 41\%             & 25\% \\
Fraction Down                & 100\%         & 36\%                & 30\%              & 63\%                & 61\%                  & 61\%           & 59\%             & 75\% \\
Fraction Tracks              & $ > 90 \%$    & 71\%                & 20\%              & 21\%                & 19\%                  & 28\%           & 26\%             & 20\% \\
Fraction Showers             & $ < 10 \%$    & 29\%                & 80\%              & 79\%                & 81\%                  & 72\%           & 74\%             & 80\% \\
\end{tabular}

\caption{Properties of events and models. Limits on the prompt flux are from \cite{ic59_muons}. The best-fit per-flavor $E^{-2}$ normalization is $E^2 \Phi_{\nu}(E) = 0.95 \cdot 10^{-8} \, \mathrm{GeV}\, \mathrm{cm}^{-2}\, \mathrm{s}^{-1}\, \mathrm{sr}^{-1}$. The global best-fit spectrum is $E^2 \phi(E) = 1.5 \times 10^{-8} (E / 100 \mathrm{TeV})^{-0.3} \mathrm{GeV} \mathrm{cm}^{-2} \mathrm{s}^{-1} \mathrm{sr}^{-1}$. As event 32, a set of coincident muons, is not reconstructable, it is excluded from all but the first row of the table. Fractions for up/down and shower/track classifications are provided at the bottom; shower/track fractions for the muon background are estimates based on examination of lower-energy events. Note that the total track rate here is dominated by the highly uncertain muon background rate. The column labeled \emph{Sum} shows the sum of all predictions given their nominal values and does not include any uncertainties in its constituent rates or the results of the best-fit background rates, which were slightly below expectations (Suppl.~Fig.~\ref{fig:llhspace}). A graphical presentation of the evolution of the up/down ratio with energy can be found in Suppl.~Fig.~\ref{fig:zenithdistslices}. The track to cascade ratio is a strong function of spectrum due to threshold effects \cite{hese_paper} that give higher efficiency in the threshold region for $\nu_e$ CC. This causes the near equality between this ratio for the $E^{-2}$ test flux and the substantially softer charm background.}
\label{tab:expectations}
\end{table*}

\section{Time Clustering Analysis}
We performed two tests for clustering of events in time, following an identical procedure to that in \cite{hese_paper}. The method is reviewed below.
The first test looked for significant time clusters in all events (neglecting, like the point source search, events 28 and 32).
A second searched for time clustering in eleven subsets of the events that formed possible spatial clusters of two or more events. 

We applied an unbinned maximum likelihood method as in \cite{ps_method} to identify timing clusters relative to an assumed constant arrival rate in both the full sample and each spatial group.
This selects the most significant time cluster over a discrete set of time windows ($\Delta t_j$), one for each possible pair of the 35 event times. Each event pair defines a start and end time ($t^{\mathrm{min}}_j$ and $t^{\mathrm{max}}_j$), with a duration $\Delta t_{j}=t_{j}^{\mathrm{max}}-t_{j}^{\mathrm{min}}$. These quantities form the signal likelihood:
\begin{equation}
S^{\mathrm{time}}_i = \frac{H(t_{\mathrm{max}}- t_j) \times H(t_j -t_{\mathrm{min}})}{t_{\mathrm{max}}-t_{\mathrm{min}}}
\end{equation}
where $H$ is the Heaviside step function.
For each $\Delta t_j$, the likelihood ratio to a flat alternative was used as a test statistic ($\mathrm{TS}_j$) as in \cite{ps_method}. 
Significance was determined by comparing the highest TS with the distribution obtained for data sets scrambled in time.  

Using all events, the most significant time cluster contained seven events (18, 19, 20, 21, 22, 23, 24). The fitted number of signal events ($\hat{n}_s$) is 6.09, with a duration of 27.3 days.  
The probability to observe a cluster this significant or better by chance is 11\%.

The second test searched for time clustering among events found in several spatial groups. Within each group, the same scrambling approach and analysis was applied as to the full sample, but with fewer events.
Results are shown in Suppl. Table~\ref{tab:times_cluster}.
The highest fluctuation observed corresponds to Cluster K with a pre-trial p-value 4.0\%.
Including trial factors due to the 11 spatial groups gives an overall post-trial  p-value for this excess of 33\%.
 
\begin{table}
\begin{tabular}{c | c c || c c c} 
	&no. of&            &                          &   \\
	&events&  event IDs &$\hat{n}_s$ & $\Delta t_{cl.}$&p-value                        \\ \hline
Cluster A & 6    & 2, 14, 22, 24, 25, 33 &2.9 & 25&17\%     \\
Cluster B & 2    & 15, 12            &2.0 &44 &9\%     \\
Cluster C & 2    & 10, 21            &2.0 &241 &38\%     \\
Cluster D & 3    & 3, 6, 27          &3.0 &558 &62\%     \\
Cluster E & 2    & 9, 26             &2.0 &294&50\%     \\
Cluster F & 2    & 16, 23            &2.0 &151 &24\%     \\
Cluster G & 2    & 8, 16             &2.0 &190 &32\%     \\
Cluster H & 3    & 19, 20, 30        &2.0 & 4  &8\%     \\
Cluster I & 2    & 4, 35             & 2.0 & 788  &94\%     \\
Cluster J & 2    & 17, 36            & 2.0 & 508  &72\%     \\
Cluster K & 3    & 29, 33, 34        & 3.0 & 120  &4\%     \
\end{tabular}
\caption{Time clustering of 11 spatially clustered event groups. All p-values are pre-trial. $\Delta t_{cl.}$, the best-fit duration, is in units of days.}
\label{tab:times_cluster}
\end{table}

\newpage
\onecolumngrid
\newpage
\section*{Event 29}

\includegraphics[width=0.8\linewidth]{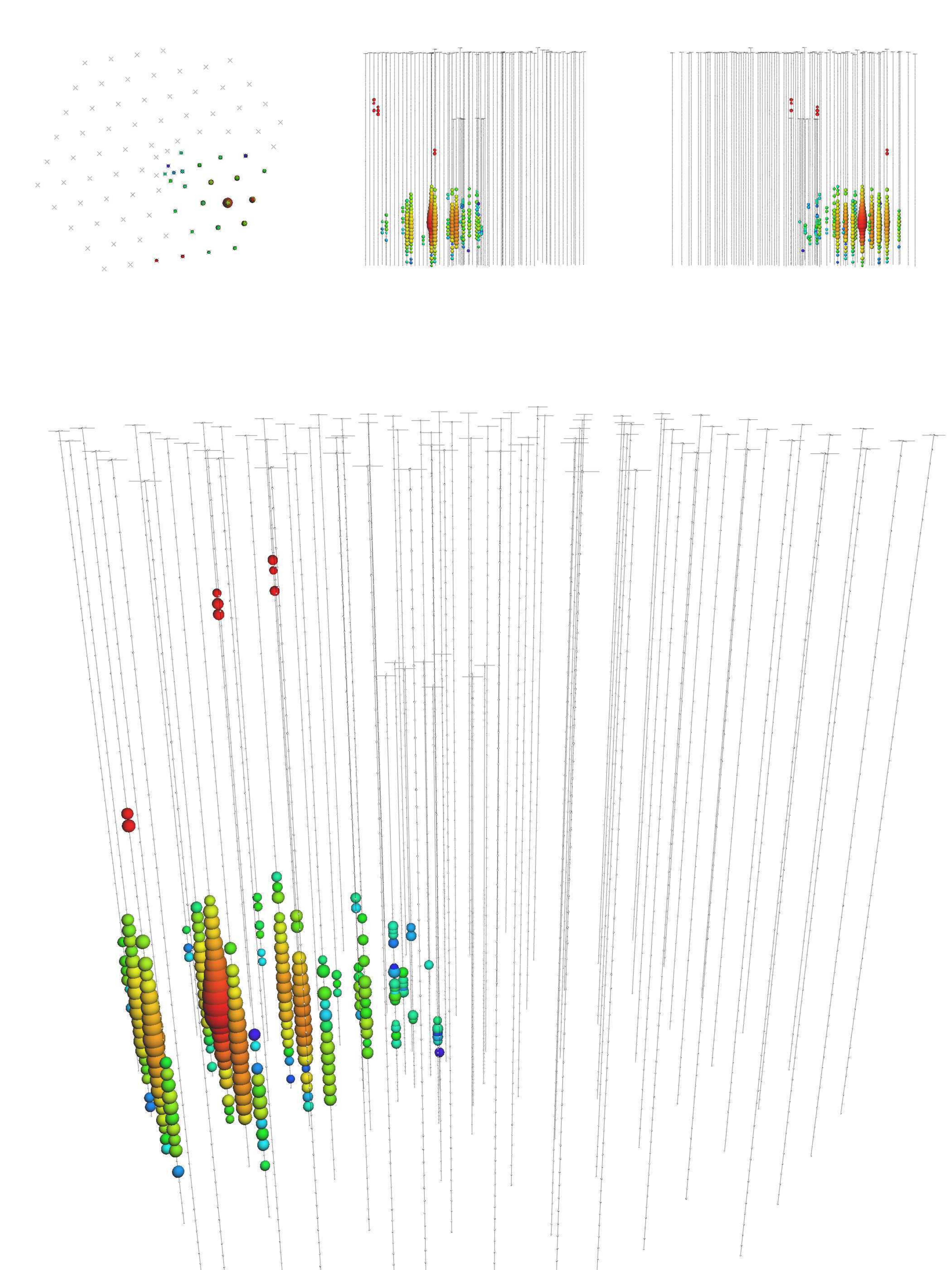}\\
\\
\includegraphics[width=0.8\linewidth]{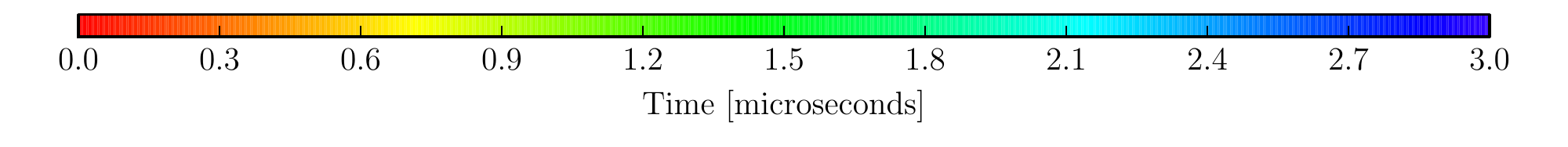}\vspace{0.2in}\\
\begin{tabular}{c|c|c|c|c|c}
Deposited Energy (TeV) & Time (MJD) & Declination (deg.) & RA (deg.) & Med. Ang. Resolution (deg.) & Topology\\
\hline
$32.7 \,^{+3.2}_{-2.9}$ & 56108.2572046 & $41.0$ & $298.1$ & $7.4$ & Shower
\end{tabular}
\newpage
\section*{Event 30}

\includegraphics[width=0.8\linewidth]{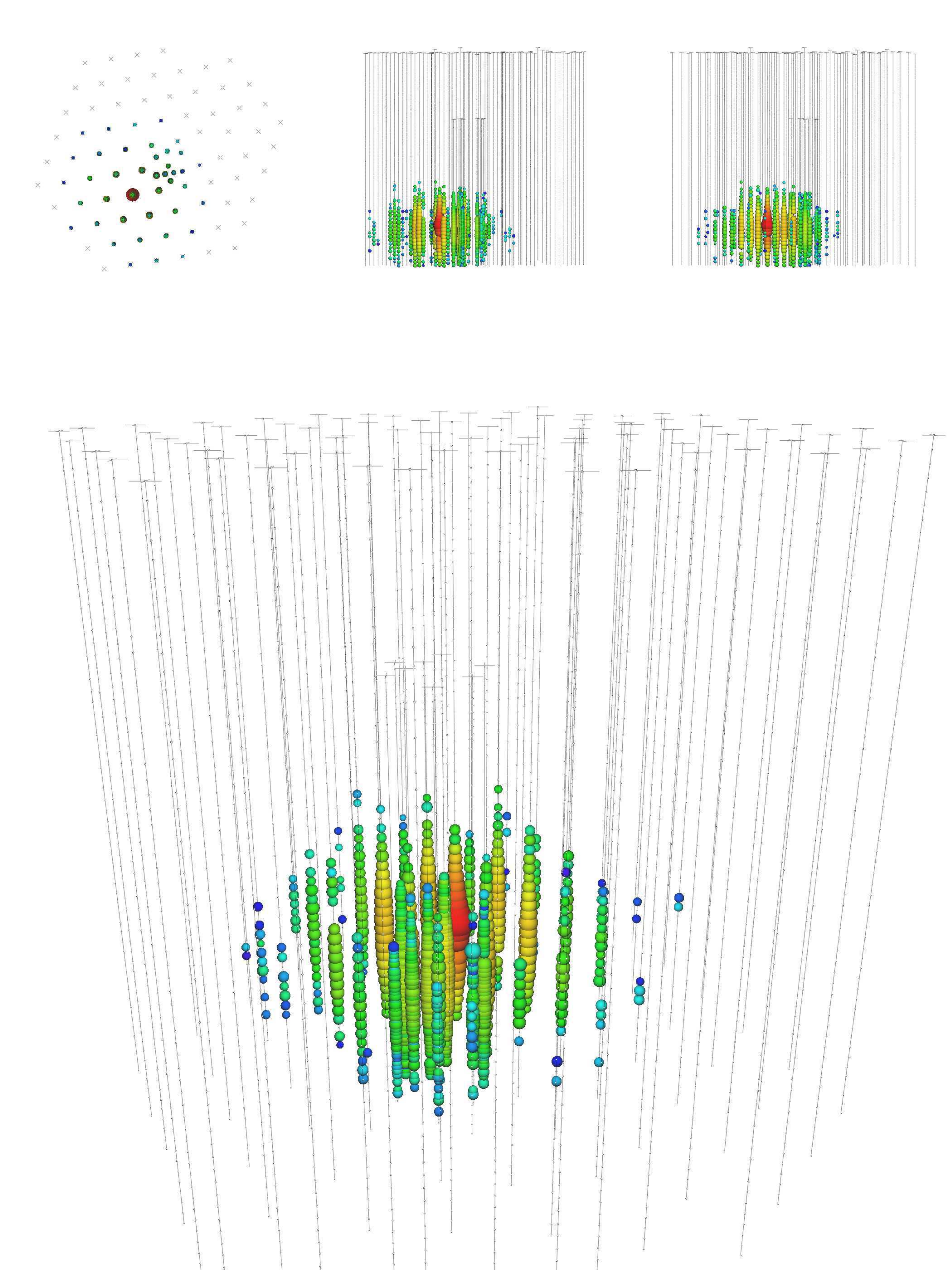}\\
\\
\includegraphics[width=0.8\linewidth]{colorbar.pdf}\vspace{0.2in}\\
\begin{tabular}{c|c|c|c|c|c}
Deposited Energy (TeV) & Time (MJD) & Declination (deg.) & RA (deg.) & Med. Ang. Resolution (deg.) & Topology\\
\hline
$129 \,^{+14}_{-12}$ & 56115.7283574 & $-82.7$ & $103.2$ & $8.0$ & Shower
\end{tabular}
\newpage
\section*{Event 31}

\includegraphics[width=0.8\linewidth]{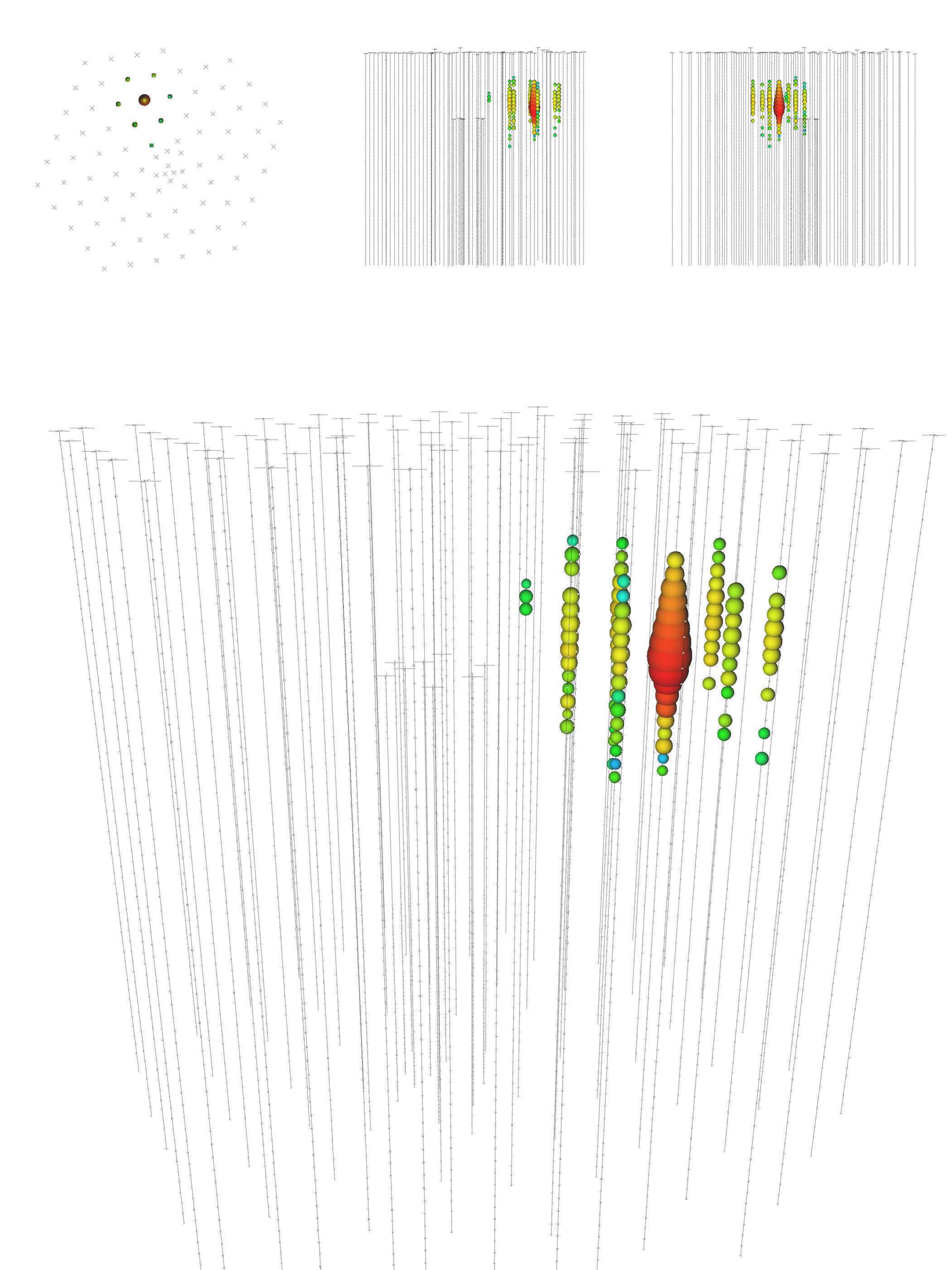}\\
\\
\includegraphics[width=0.8\linewidth]{colorbar.pdf}\vspace{0.2in}\\
\begin{tabular}{c|c|c|c|c|c}
Deposited Energy (TeV) & Time (MJD) & Declination (deg.) & RA (deg.) & Med. Ang. Resolution (deg.) & Topology\\
\hline
$42.5 \,^{+5.4}_{-5.7}$ & 56176.3914143 & $78.3$ & $146.1$ & $26.0$ & Shower
\end{tabular}
\newpage
\section*{Event 32}

\includegraphics[width=0.8\linewidth]{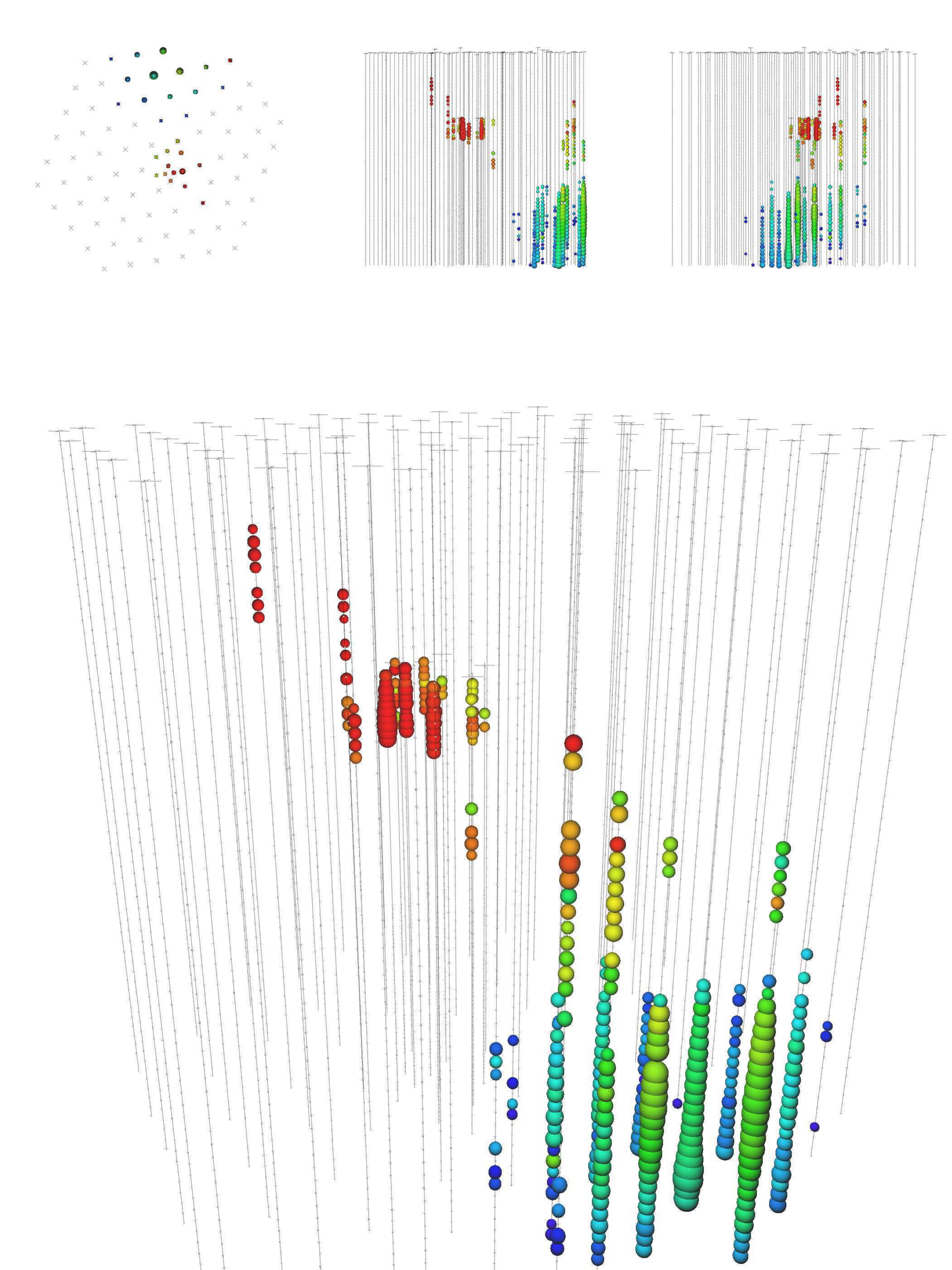}\\
\\
\includegraphics[width=0.8\linewidth]{colorbar.pdf}\vspace{0.2in}\\
\begin{tabular}{c|c|c|c|c|c}
Deposited Energy (TeV) & Time (MJD) & Declination (deg.) & RA (deg.) & Med. Ang. Resolution (deg.) & Topology\\
\hline
--- & 56211.7401231 & --- & --- & --- & Coincident
\end{tabular}
\newpage
\section*{Event 33}

\includegraphics[width=0.8\linewidth]{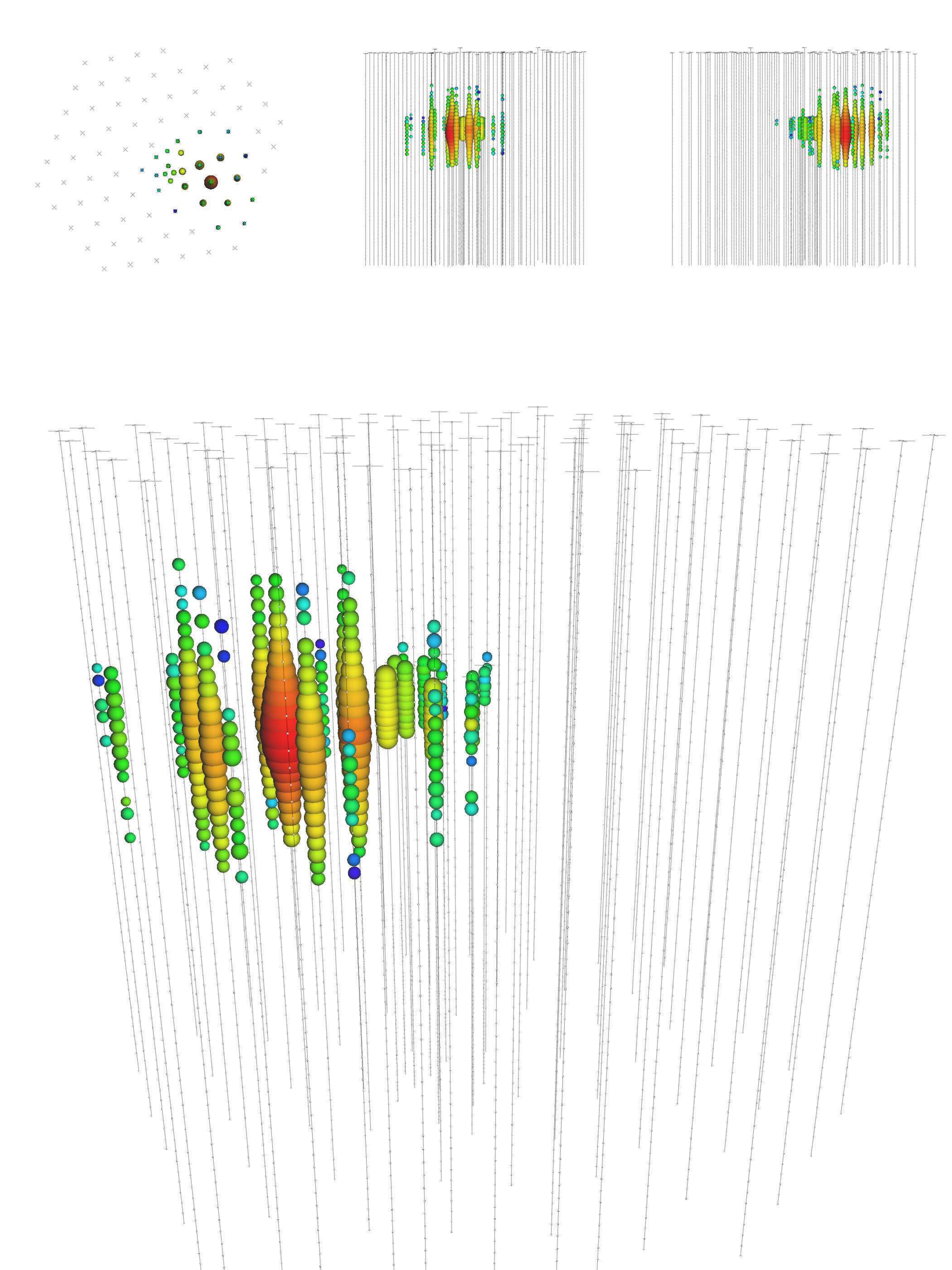}\\
\\
\includegraphics[width=0.8\linewidth]{colorbar.pdf}\vspace{0.2in}\\
\begin{tabular}{c|c|c|c|c|c}
Deposited Energy (TeV) & Time (MJD) & Declination (deg.) & RA (deg.) & Med. Ang. Resolution (deg.) & Topology\\
\hline
$385 \,^{+46}_{-49}$ & 56221.3424023 & $7.8$ & $292.5$ & $13.5$ & Shower
\end{tabular}
\newpage
\section*{Event 34}

\includegraphics[width=0.8\linewidth]{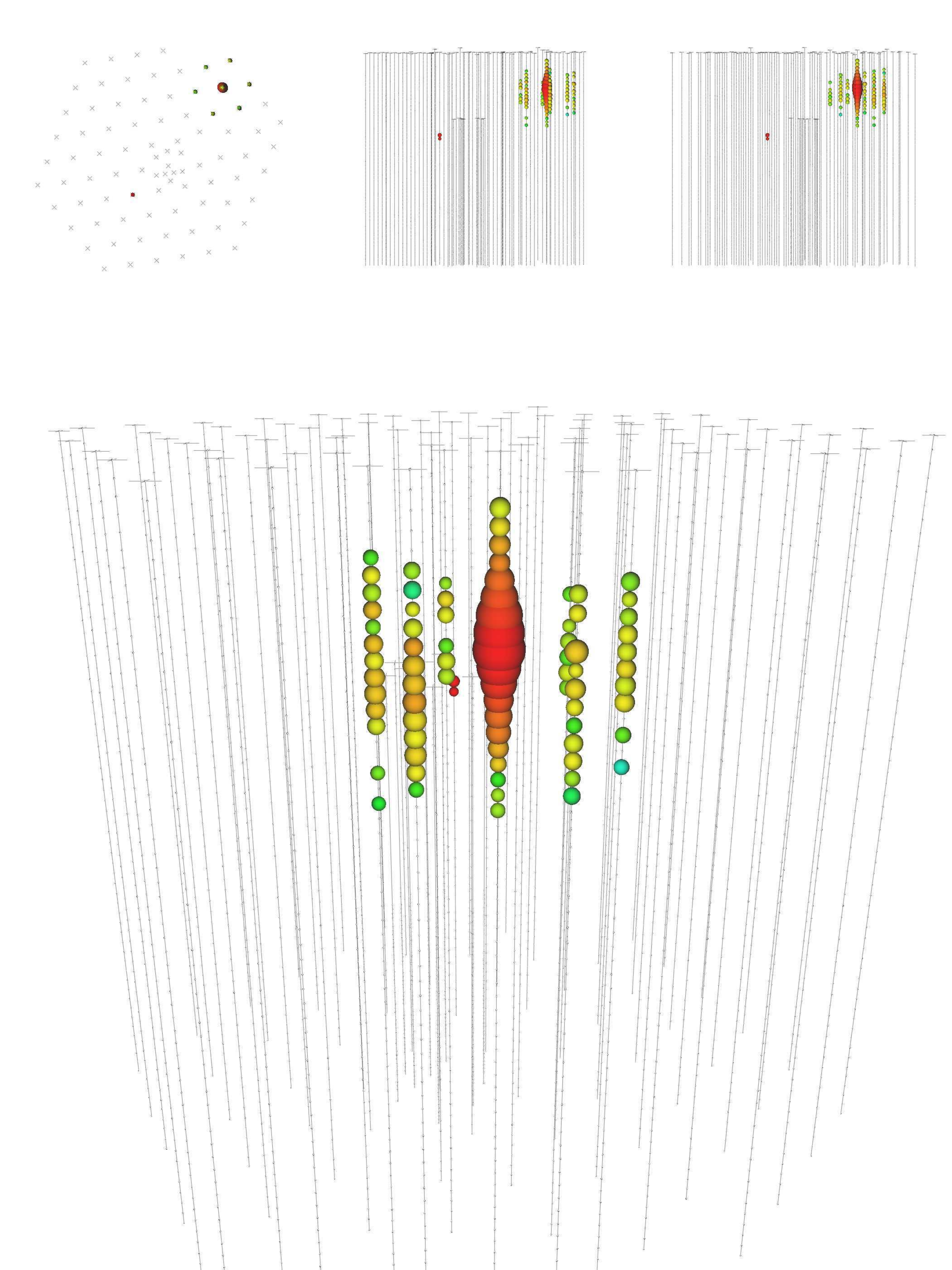}\\
\\
\includegraphics[width=0.8\linewidth]{colorbar.pdf}\vspace{0.2in}\\
\begin{tabular}{c|c|c|c|c|c}
Deposited Energy (TeV) & Time (MJD) & Declination (deg.) & RA (deg.) & Med. Ang. Resolution (deg.) & Topology\\
\hline
$42.1 \,^{+6.5}_{-6.3}$ & 56228.6055226 & $31.3$ & $323.4$ & $42.7$ & Shower
\end{tabular}
\newpage
\section*{Event 35}

\includegraphics[width=0.8\linewidth]{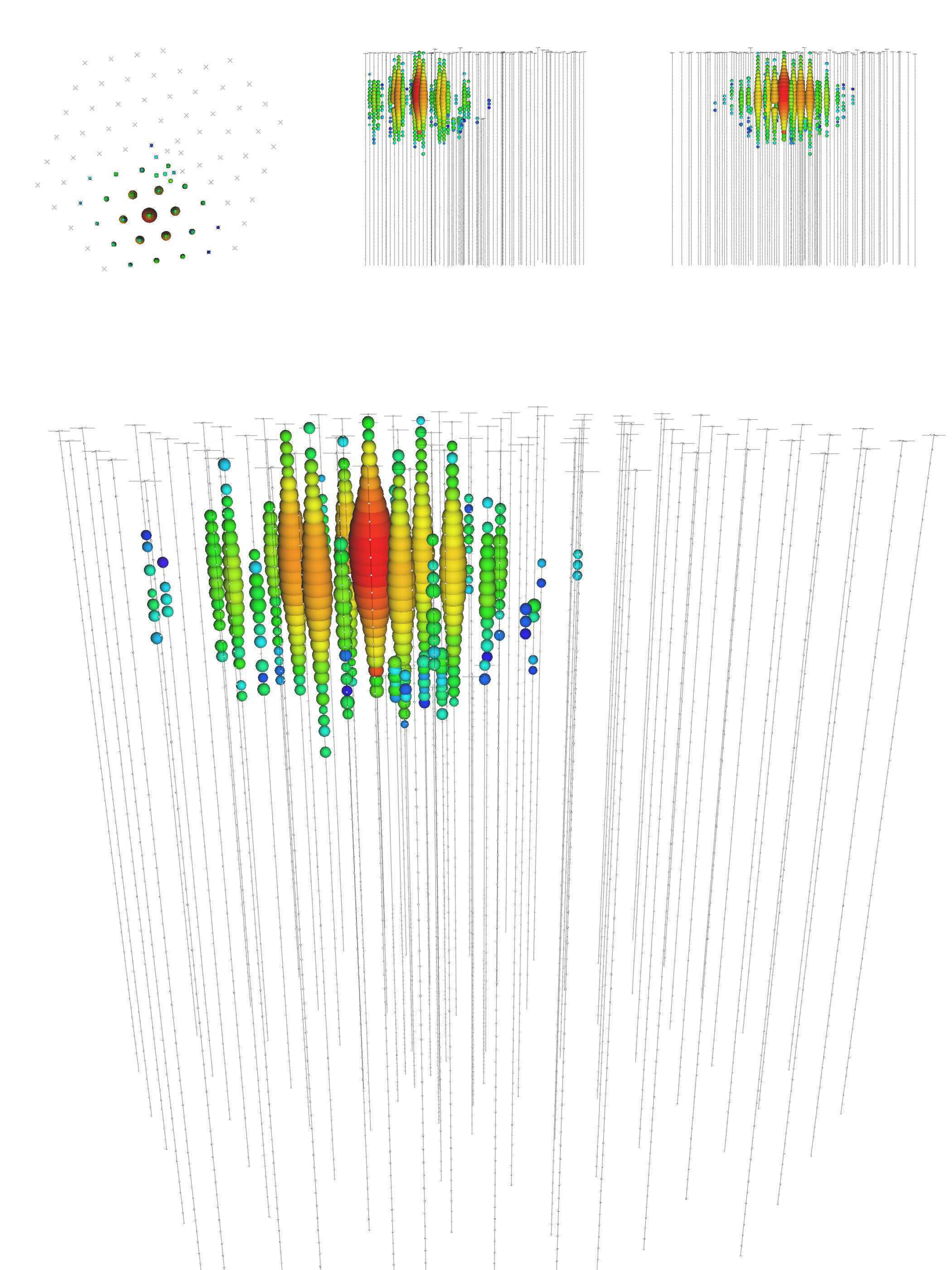}\\
\\
\includegraphics[width=0.8\linewidth]{colorbar.pdf}\vspace{0.2in}\\
\begin{tabular}{c|c|c|c|c|c}
Deposited Energy (TeV) & Time (MJD) & Declination (deg.) & RA (deg.) & Med. Ang. Resolution (deg.) & Topology\\
\hline
$2004 \,^{+236}_{-262}$ & 56265.1338677 & $-55.8$ & $208.4$ & $15.9$ & Shower
\end{tabular}
\newpage
\section*{Event 36}

\includegraphics[width=0.8\linewidth]{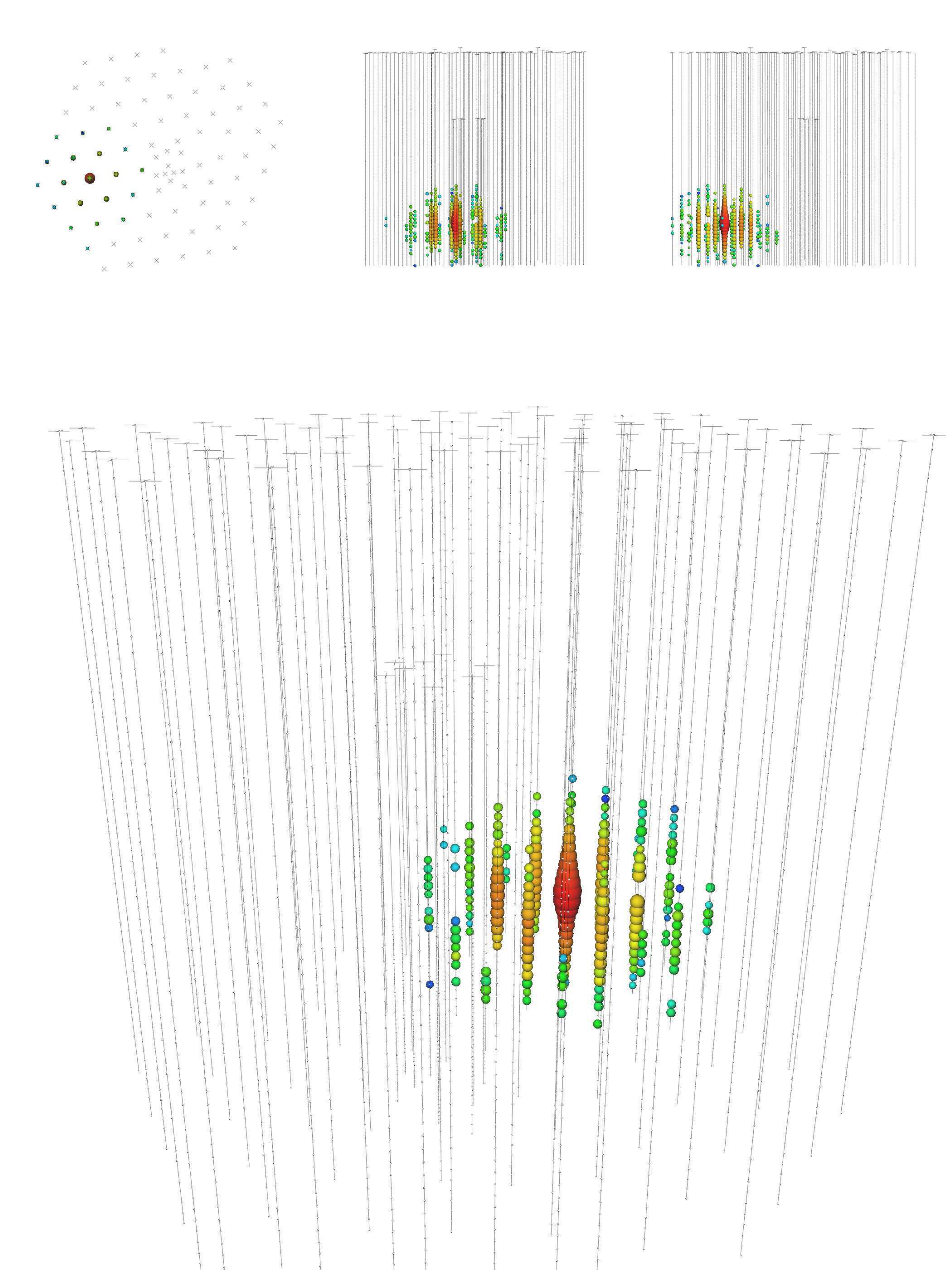}\\
\\
\includegraphics[width=0.8\linewidth]{colorbar.pdf}\vspace{0.2in}\\
\begin{tabular}{c|c|c|c|c|c}
Deposited Energy (TeV) & Time (MJD) & Declination (deg.) & RA (deg.) & Med. Ang. Resolution (deg.) & Topology\\
\hline
$28.9 \,^{+3.0}_{-2.6}$ & 56308.1642740 & $-3.0$ & $257.7$ & $11.7$ & Shower
\end{tabular}
\newpage
\section*{Event 37}

\includegraphics[width=0.8\linewidth]{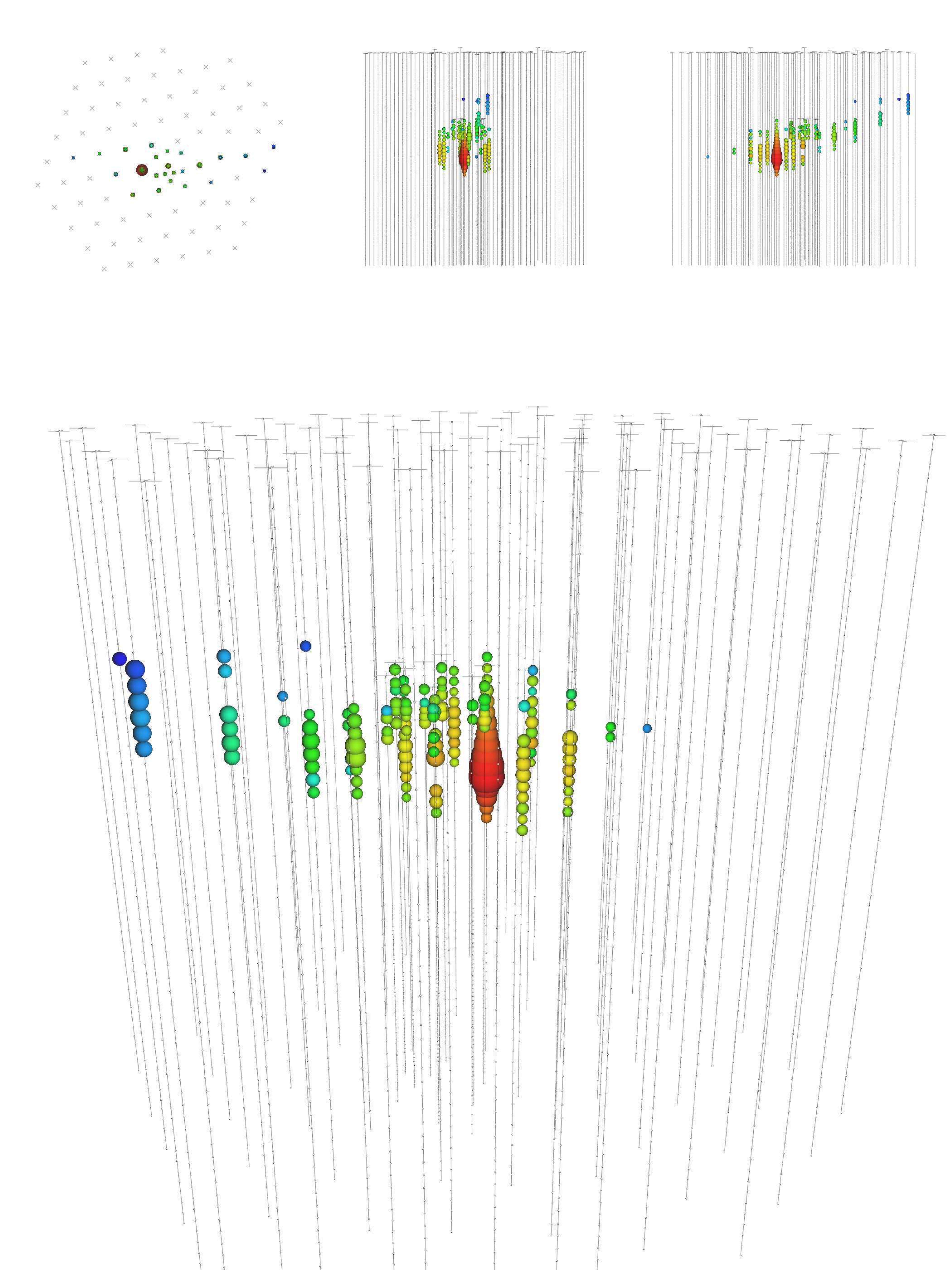}\\
\\
\includegraphics[width=0.8\linewidth]{colorbar.pdf}\vspace{0.2in}\\
\begin{tabular}{c|c|c|c|c|c}
Deposited Energy (TeV) & Time (MJD) & Declination (deg.) & RA (deg.) & Med. Ang. Resolution (deg.) & Topology\\
\hline
$30.8 \,^{+3.3}_{-3.5}$ & 56390.1887627 & $20.7$ & $167.3$ & $\lesssim 1.2$ & Track
\end{tabular}

\clearpage

\end{document}